\documentclass{svjour3}
\usepackage{graphicx}

\begin{document}

\title{Galactic tide and orbital evolution of comets}
\author{L. K\'{o}mar \and J. Kla\v{c}ka \and P. P\'{a}stor}
\institute{Faculty of Mathematics,
Physics and Informatics, Comenius University \\
Mlynsk\'{a} dolina, 842 48 Bratislava, Slovak Republic \\
\email{\{komar,klacka,pavol.pastor\}@fmph.uniba.sk}}

\date{}

\authorrunning{L. K\'{o}mar et al.}
\titlerunning{Motion of dust in mean motion resonances with planets}
\maketitle

\begin{abstract}
Equation of motion for a comet in the Oort cloud is numerically solved.
Orbital evolution of the comet under the action of the gravity of the Sun and the Galaxy
is presented for various initial conditions. 

Oscillations of the Sun with respect to the galactic equatorial plane
are taken into account. Real values of physical quantities concerning the gravitational action of the 
galactic neighbourhood of the Sun are important. The results are compared with currently used 
more simple models of the galactic tide. It turns out that physically improved 
models yield results which significantly differ from the results obtained on 
the basis of the conventional models. E.g., the number of returns of the comets into the inner 
part of the Solar System are about two times greater than it is in the conventional  models.

It seems that a comet from the Oort cloud can be a source of the dinosaurs extinction
at about 65 Myr ago. A close encounter of a star or an interstellar cloud disturbed a comet of the Oort 
cloud in the way that its semi-major axis increased/decreased above the value  
5 $\times$ 10$^{4}$ AU and the comet hit the Earth.

\keywords{Oort cloud \and Galaxy \and Comets \and Equation of motion \and
Orbital evolution}
\end{abstract}

\section{Introduction}
Global galactic gravitational field influences motion of a comet in the
\"{O}pik-Oort cloud (\"{O}pik 1932, Oort 1950) in the form of the galactic tide.
The motion of the comet with respect
to the Sun is important in better understanding of the Oort cloud.
This paper deals with detailed numerical solution of the equation of motion
of the comet under the gravity of the Sun and the galactic tide.
We consider equation of motion presented by Kla\v{c}ka (2009a, 2009b).
The $x-$ and $y-$ components of the acceleration come not only
from the $x-$ and $y-$ components of the position of the comet, but also from
the $z-$component of the position due to the gravity of the galactic disk.
Our paper provides detailed numerical solutions of the equation of motion.
The importance of the $\Gamma$-terms is tested. Comparison with
the conventional models of galactic tide is discussed (see, e.g., Mihalas 1968,
Heilser and Tremaine 1986, Levison et al. 2001, Dybczynski et al. 2008).

\section{Orbital elements}
We discuss several models of galactic tide.
Each of them is represented by an equation of motion.
The equation of motion is numerically solved. On the basis of the positional
and velocity vectors at each position, we can find orbital
elements and their evolution. We can use Eq. (47) in Kla\v{c}ka (2004)
for this purpose (the right-hand side of the last equation in Eq. 47
contains $1/e$ instead of 1). We can summarize the equations in the
following form [osculating orbital elements: $a$ -- semi-major axis;
$e$ -- eccentricity; $i$ -- inclination of the orbital plane to the
reference plane -- galactic equatorial plane;
$\Omega$ -- longitude of the ascending node; $\omega$ --
longitude of pericenter (the argument of pericenter/perihelion); 
$q$ -- perihelion distance; $Q$ -- aphelion distance; $\Theta$
is the position angle of the particle on the orbit, when measured
from the ascending node in the direction of the particle's motion,
$\Theta = \omega + f$]:
\begin{eqnarray}\label{1}
\vec{r} &=& ( \xi, \eta, \zeta ) ~, ~~ r = | \vec{r} | ~,
\nonumber \\
\vec{v} &\equiv& \dot{\vec{r}} = ( \dot{\xi}, \dot{\eta}, \dot{\zeta} ) ~,
\nonumber \\
E &=& \frac{1}{2} ~\vec{v}^{2} ~-~ \frac{G ~M_{\odot}}{r} ~,
\nonumber \\
\vec{H} &=& \vec{r} \times \vec{v} ~,
\nonumber \\
p &=& \frac{\vec{H}^{2}}{G ~M_{\odot}} ~,
\nonumber \\
e &=& \sqrt{1 ~+~ 2 ~\frac{p~ E}{G ~M_{\odot}}} ~,
\nonumber \\
a &=& \frac{p}{1 - e^{2}} ~,
\nonumber \\
E &=& -~ \frac{G~ M_{\odot}}{2~p} ~ \left ( 1 - e^{2} \right ) ~,
\nonumber \\
q &=& a ( 1 - e ) ~,
\nonumber \\
Q &=& a (1 + e)~,
\nonumber \\
i &=& \arccos{ \left ( \frac{H_{z}}{| \vec{H} |} \right ) } ~,
\nonumber \\
\sin \Omega &=& \frac{H_{\xi}}{ \sqrt{H_{\xi}^{2} + H_{\eta}^{2}} } ~,
\nonumber \\
\cos \Omega &=& -~ \frac{H_{\eta}}{ \sqrt{H_{\xi}^{2} + H_{\eta}^{2}} } ~,
\nonumber \\
\sin \omega &=& \frac{| \vec{H} |}{ \sqrt{H_{\xi}^{2} + H_{\eta}^{2}} } ~
        \frac{1}{r ~e} ~\times~ S1
\nonumber \\
S1 &=& -~ \frac{\eta~H_{\xi} ~-~\xi~H_{\eta}}{| \vec{H} |} ~
       \frac{\vec{v} \cdot \vec{e}_{R}}{\sqrt{G ~M_{\odot} / p}} ~+~
     \zeta ~\left ( \frac{\vec{v} \cdot \vec{e}_{T}}{\sqrt{G ~M_{\odot} / p}}
     ~-~ 1 \right ) ~,
\nonumber \\
\cos \omega &=& \frac{| \vec{H} |}{ \sqrt{H_{\xi}^{2} + H_{\eta}^{2}} } ~
        \frac{1}{r ~e} ~\times~ C1
\nonumber \\
C1 &=&  \frac{\eta~H_{\xi} ~-~ \xi~H_{\eta}}{| \vec{H} |} ~
    \left ( \frac{\vec{v} \cdot \vec{e}_{T}}{\sqrt{G ~M_{\odot} / p}}
    ~-~ 1 \right ) ~+~ \zeta ~
       \frac{\vec{v} \cdot \vec{e}_{R}}{\sqrt{G ~M_{\odot} / p}} ~,
\nonumber \\
\vec{e}_{R} &=& \frac{\vec{r}}{r} ~,
\nonumber \\
\vec{e}_{N} &=& \frac{\vec{H}}{| \vec{H} |} ~,
\nonumber \\
\vec{e}_{T} &=& \vec{e}_{N} \times \vec{e}_{R} ~.
\end{eqnarray}

\section{Current models}

\subsection{Simple model}

The most simple model considering galactic tide is described by the well-known equation
of motion (see, e. g., Mihalas 1968, Heisler and Tremaine 1986)
\begin{eqnarray}\label{2}
\frac{d^{2} \xi}{dt^{2}} &=& - ~ \frac{G M_{\odot}}{r^{3}} ~ \xi
\nonumber \\
\frac{d^{2} \eta}{dt^{2}} &=& - ~ \frac{G M_{\odot}}{r^{3}} ~ \eta
 \nonumber \\
\frac{d^{2} \zeta}{dt^{2}} &=& - ~ \frac{G M_{\odot}}{r^{3}} ~ \zeta
~-~ \left [ 4 ~\pi ~G ~\varrho ~+~
2 \left ( A^{2} ~-~ B^{2} \right ) \right ] ~\zeta
\nonumber \\
r &=& \sqrt{\xi ^{2} ~+~ \eta ^{2} ~+~ \zeta ^{2}} ~,
\end{eqnarray}
where $G$ is the gravitational constant, $M_{\odot}$ is mass of the Sun, 
$\rho$ is the mass density of the Galaxy and $A$, $B$ are Oort constants.
Numerical solution of the equation of motion was presented by Pretka and
Dybczynski  (1994) and secular evolution by Kla\v{c}ka and Gajdo\v{s}\'{\i}k
(2001):
\begin{eqnarray}\label{3}
 \left \langle \frac{da}{dt}\right \rangle &=& 0
\nonumber \\
 \left\langle\frac{de}{dt}\right\rangle &=& \frac{5}{4}~k~\sqrt{\frac{a^{3}}{\mu}}~(\sin~i)^{2}~[\sin(2\omega)]~e~\sqrt{1~-~e^{2}}
 \nonumber \\
 \left\langle\frac{di}{dt}\right\rangle &=& \frac{1}{2}[\sin(2\omega)](\sin~i)~\frac{(1-e_{0}^{2})~{(\cos~i_{0})}^{2}~-~(\cos~i)^{2}}{\sqrt{1~-~e_{0}^{2}}~\cos~i_{0}}
 \nonumber \\
 \left\langle\frac{d\omega}{dt}\right\rangle &=& \frac{[1/5~-~(\sin~\omega)^{2}](1~-~e_{0}^{2})(\cos~i_{0})^{2}~+~(\sin~\omega)^{2}(\cos~i)^{4}}{\sqrt{1~-~e_{0}^{2}}~(\cos~i_{0})(\cos~i)}~,
\nonumber \\
k &=&  4 ~\pi ~G ~\varrho ~+~ 2 \left ( A^{2} ~-~ B^{2} \right )  ~.
\end{eqnarray}

\subsection{Standard model}
Previous section presented the most simple action of galactic tide.
A more elaborated model is often used (Heisler and Tremaine 1986,
Levison et al. 2001,  Dybczynski et al. 2008):
\begin{eqnarray}\label{4}
\frac{d^{2} \xi}{dt^{2}} &=& - ~ \frac{G M_{\odot}}{r^{3}} ~ \xi
~+~ ( A - B ) \left [ A + B + 2 A \cos \left ( 2 ~ \omega_{0} t \right )
 \right ] ~ \xi
\nonumber \\
& & -~ 2 A ( A - B ) \sin \left ( 2 ~ \omega_{0} t \right ) ~\eta
\nonumber \\
\frac{d^{2} \eta}{dt^{2}} &=& - ~ \frac{G M_{\odot}}{r^{3}} ~ \eta
~-~  2 A ( A - B ) \sin \left ( 2 ~ \omega_{0} t \right ) ~ \xi
\nonumber \\
& & +~ ( A - B ) \left [ A + B - 2 A \cos \left ( 2 ~ \omega_{0} t \right )
 \right ] ~ \eta
 \nonumber \\
\frac{d^{2} \zeta}{dt^{2}} &=& - ~ \frac{G M_{\odot}}{r^{3}} ~ \zeta
~-~ \left [ 4 ~\pi ~G ~\varrho ~+~
2 \left ( A^{2} ~-~ B^{2} \right ) \right ] ~\zeta
\nonumber \\
r &=& \sqrt{\xi ^{2} ~+~ \eta ^{2} ~+~ \zeta ^{2}} ~,
\nonumber \\
\omega_{0} &=& A ~-~ B ~,
\end{eqnarray}
where $G$ is the gravitational constant, $M_{\odot}$ is the mass
of the Sun and the numerical values of the other relevant quantities are
\begin{eqnarray}\label{5}
A &=& 13.0 ~ \mbox{km} ~\mbox{s}^{-1} ~\mbox{kpc}^{-1} ~,
\nonumber \\
B &=& -~ 13.0 ~ \mbox{km} ~\mbox{s}^{-1} ~\mbox{kpc}^{-1} ~,
\nonumber \\
\varrho &=& 0.10 ~\mbox{M}_{\odot}~ \mbox{pc}^{-3} ~.
\end{eqnarray}

\begin{figure}[h]
\centering
\includegraphics[scale=0.52]{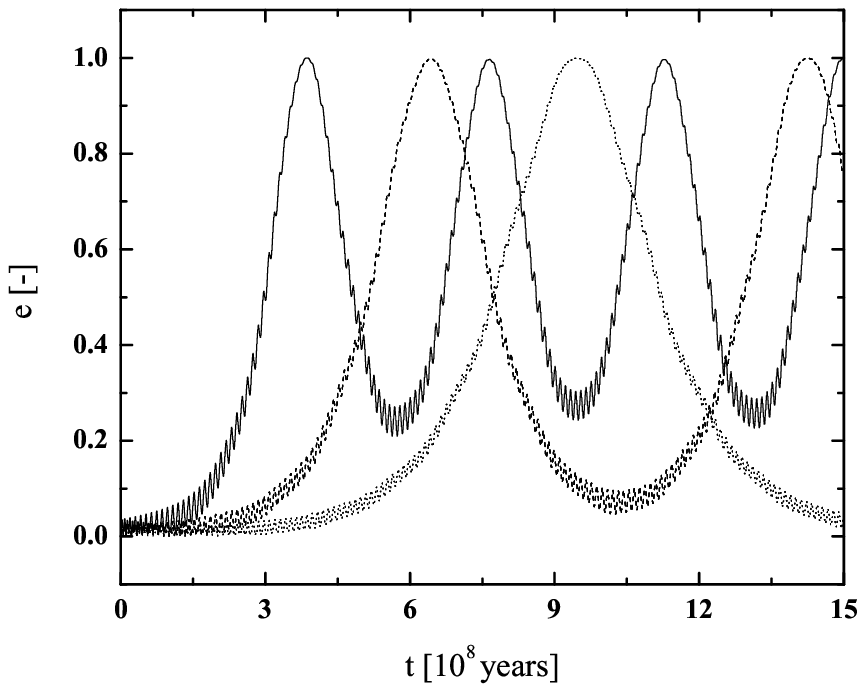}
\includegraphics[scale=0.52]{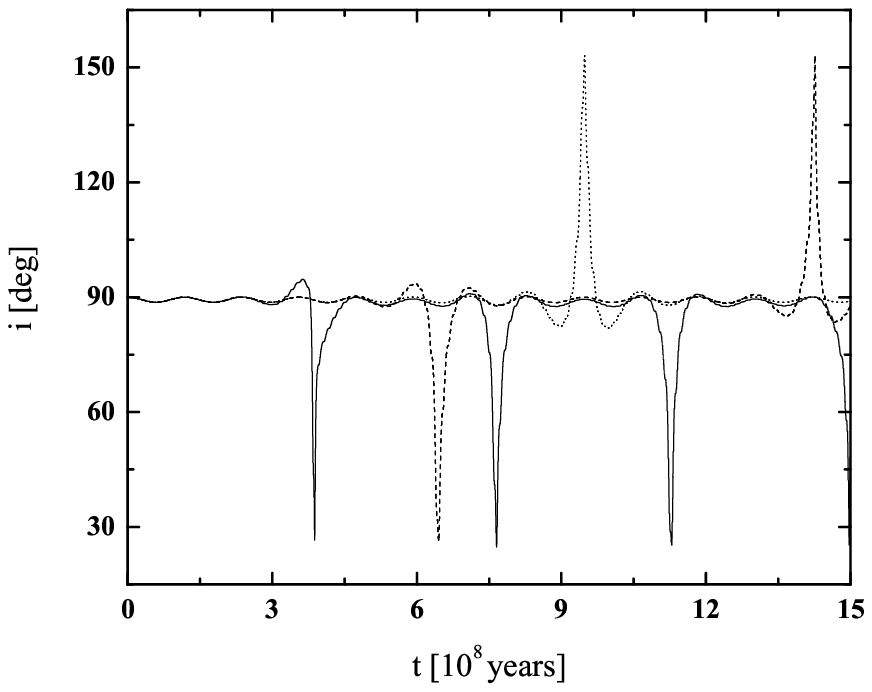}
\includegraphics[scale=0.52]{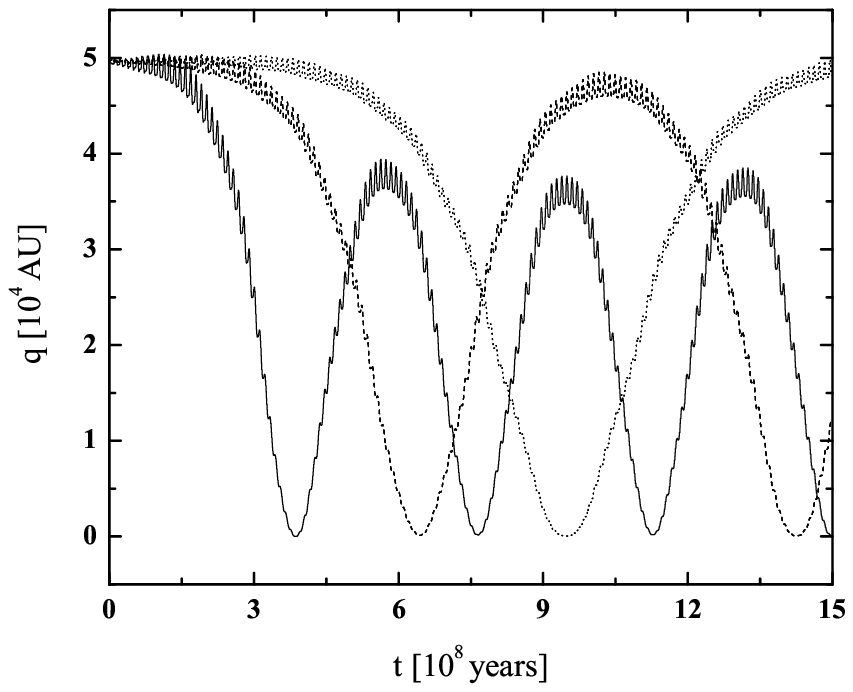}
\includegraphics[scale=0.52]{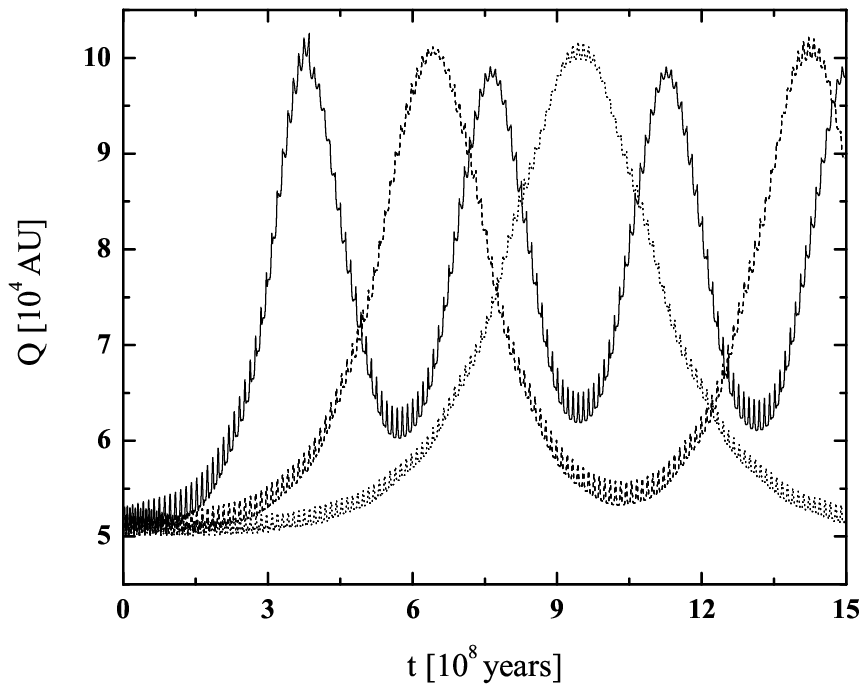}
\label{F1}
\caption{Orbital evolution of the comet situated in the Oort cloud with $a_{in}$ = $5 \times 10^{4}$ AU, $e_{in} \approx$ 0, $i_{in}$ = $90^{\circ}$
under the influence of the solar gravity and the galactic tide based on the standard model for various mass densities 
in the neighbourhood of the Sun: $\rho$  = 0.075 M$_{\odot}$ pc$^{-3}$ (dotted line), $\rho$  = 0.1 M$_{\odot}$ pc$^{-3}$ (dashed line), 
$\rho$  = 0.15 M$_{\odot}$ pc$^{-3}$ (solid line).}
\end{figure}

Fig. 1 represents the numerical solution of Eqs. (4) and (5). It shows the orbital evolution of the comet situated in the Oort cloud 
of comets ($a_{in}$ = $5 \times 10^{4}$ AU) with initial almost circular orbit perpendicular to the plane of the galactic equator ($e_{in} \approx$ 0, $i_{in}$ = $90^{\circ}$).
Solar gravity and the galactic tide in terms of the standard model are considered for various local densities $\rho$ in the neighbourhood of the Sun.

In the literature the values of the local density span from $\rho$ = 0.075 M$_{\odot}$ pc$^{-3}$ (Cr\'{e}z\'{e} et al. 1998) to $\rho$ = 0.185 M$_{\odot}$ pc$^{-3}$ 
(Bahcall 1984). As it is shown in Fig. 1, the orbital evolution of the comet initially situated in the Oort cloud is very sensitive to the local density. For comparison 
we used the values $\rho$ = 0.075 M$_{\odot}$ pc$^{-3}$, $\rho$ = 0.1 M$_{\odot}$ pc$^{-3}$ (Levison et al. 2001) and 
$\rho$ = 0.15 M$_{\odot}$ pc$^{-3}$ (Wiegert and Tremaine 1999). 

\subsection{Comparison of the current models}

\begin{figure}[h]
\centering
\includegraphics[scale=0.52]{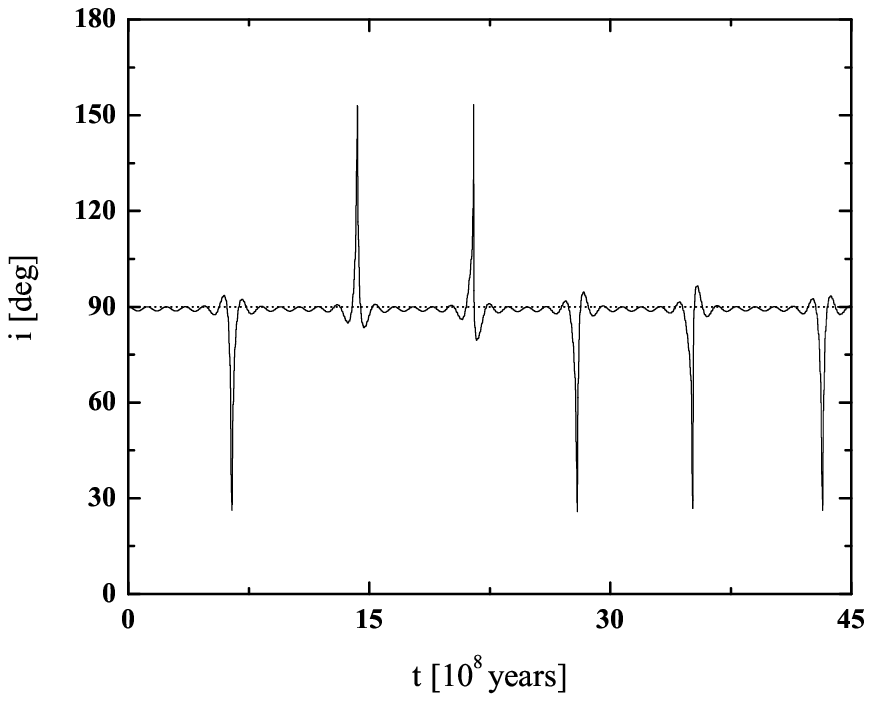}
\includegraphics[scale=0.52]{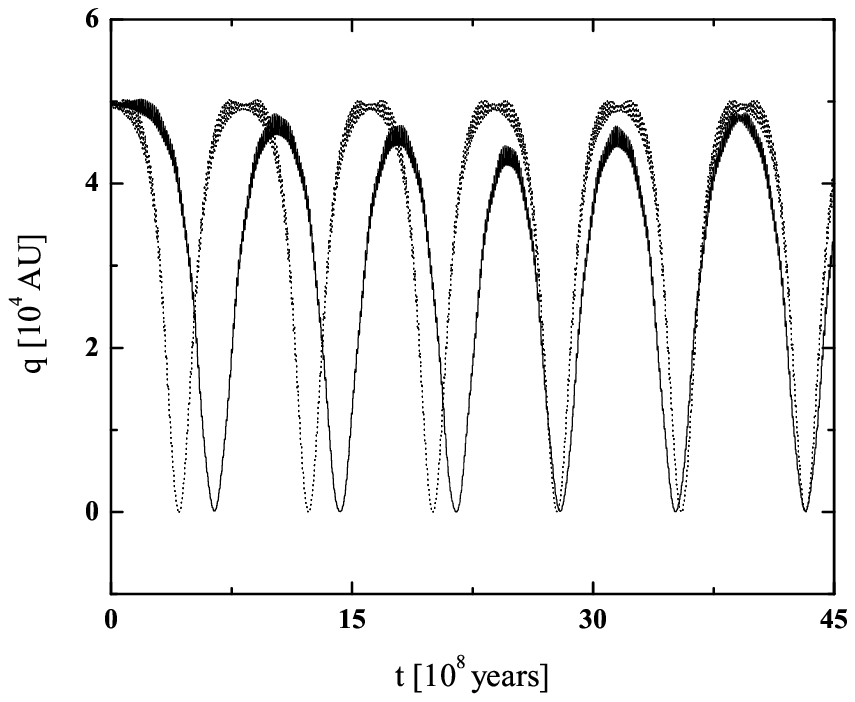}
\includegraphics[scale=0.52]{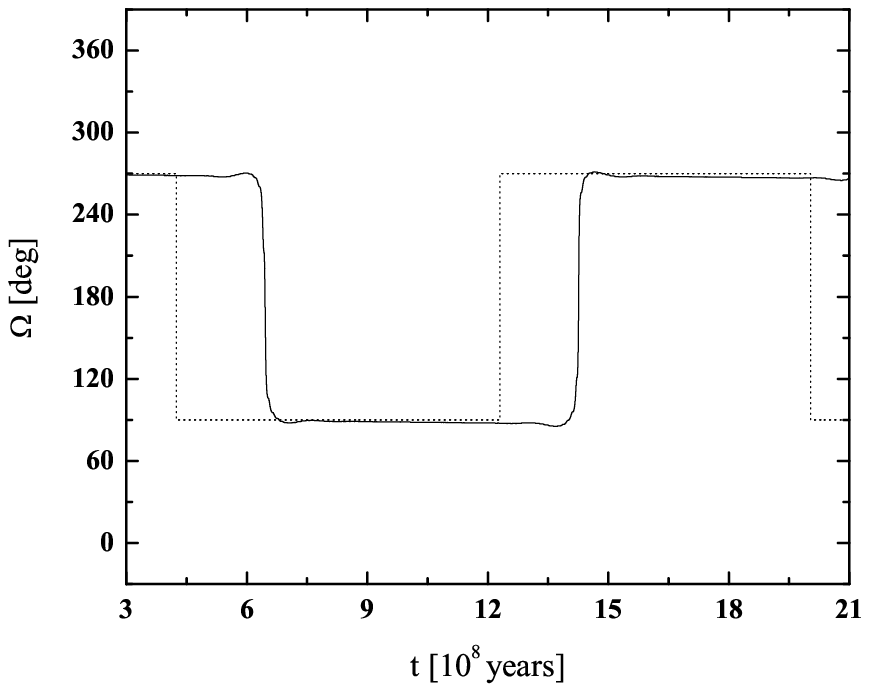}
\includegraphics[scale=0.52]{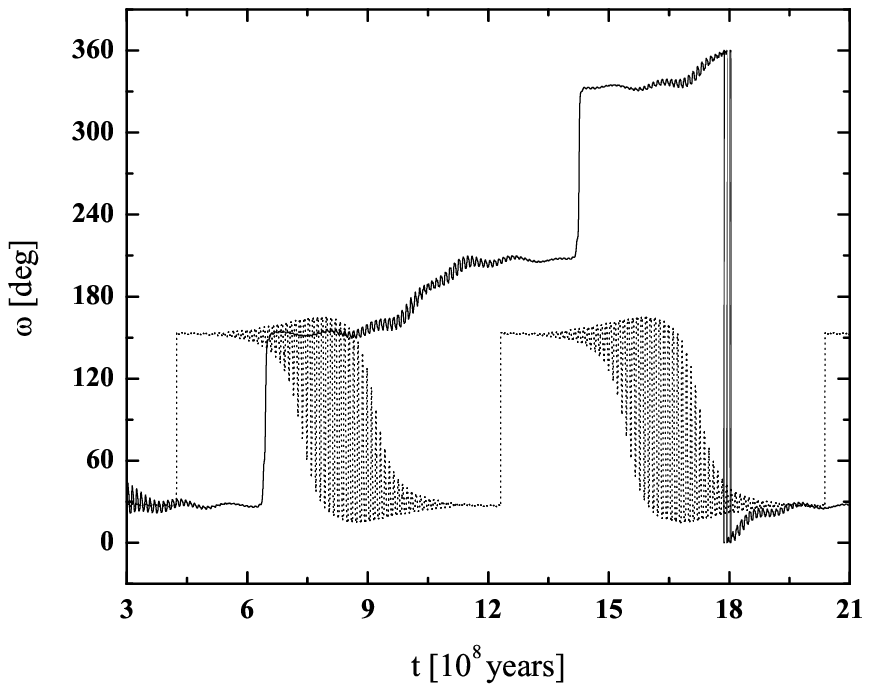}
\label{F2}
\caption{Orbital evolution of the comet situated in the Oort cloud with 
$a_{in}$ = $5 \times 10^{4}$ AU, $e_{in} \approx 0$, $i_{in}$ = $90^{\circ}$ under the influence of the solar gravity 
and the galactic tide for the simple (dotted line) and the standard (solid line) models. 
Local density $\rho$ = 0.1 M$_{\odot}$ pc$^{-3}$ is considered for both models.}
\end{figure}

\begin{figure}[h]
\centering
\includegraphics[scale=0.52]{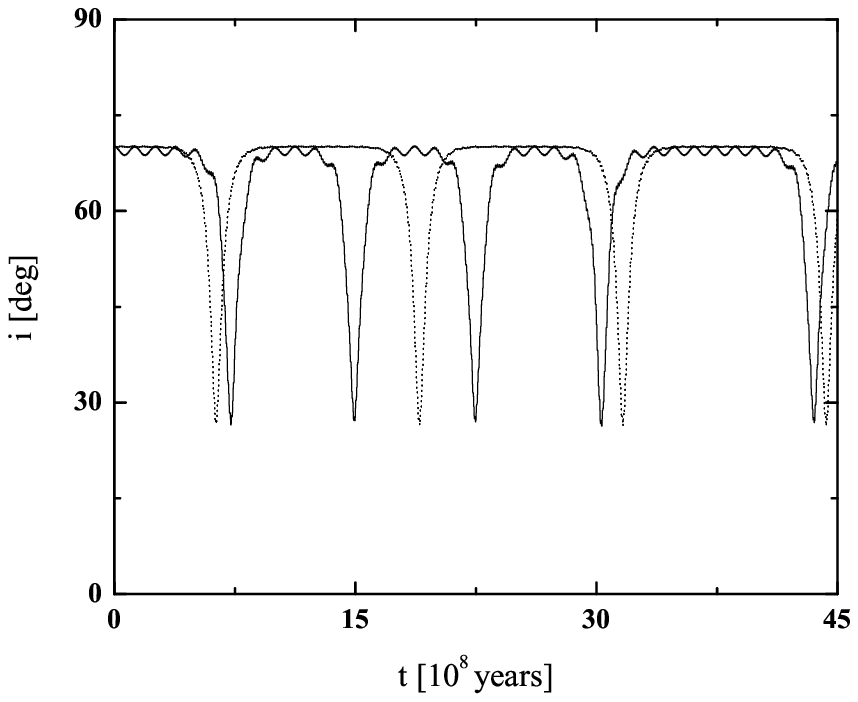}
\includegraphics[scale=0.52]{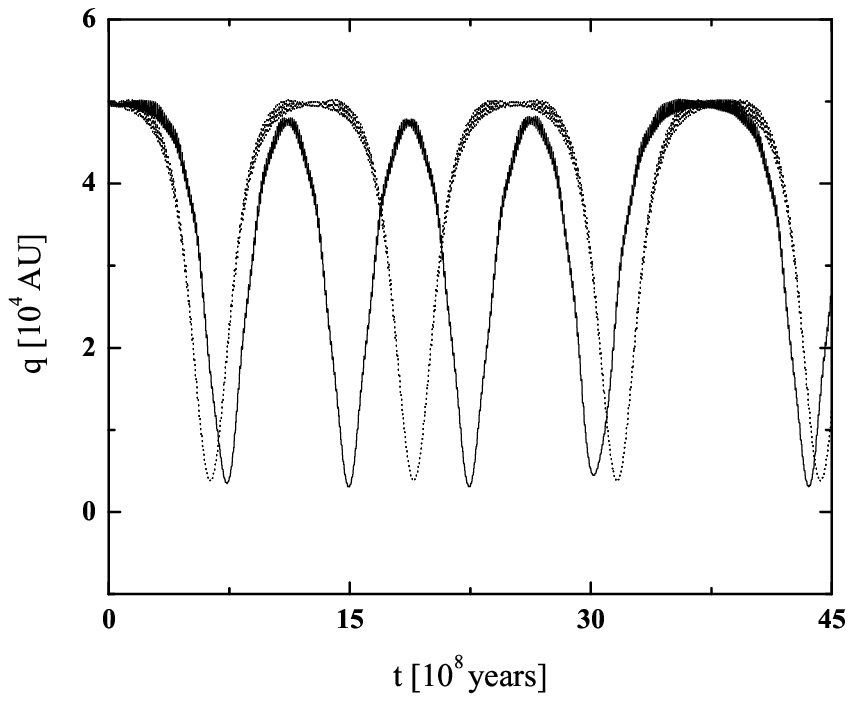} 
\includegraphics[scale=0.52]{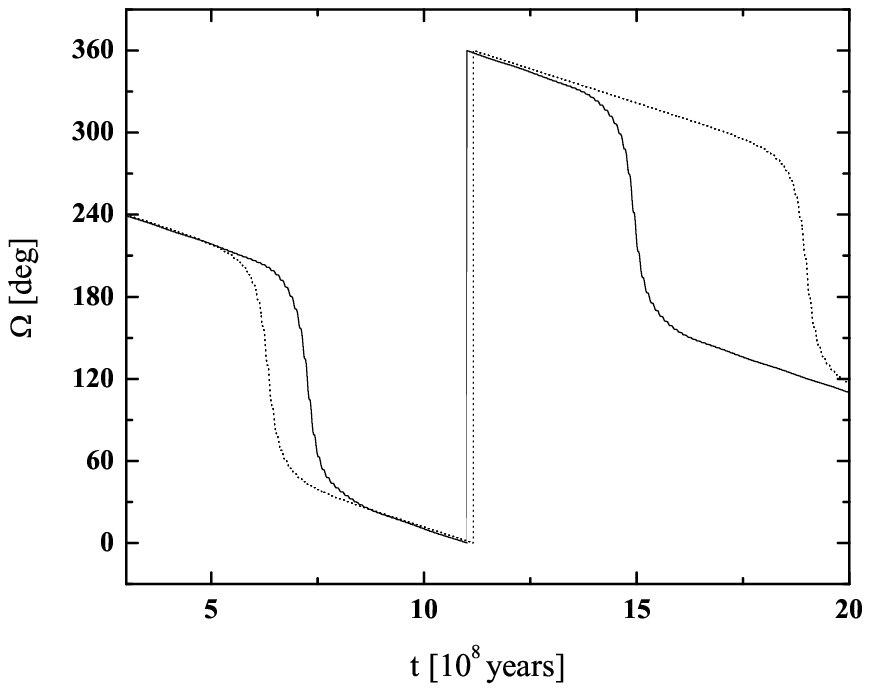}
\includegraphics[scale=0.52]{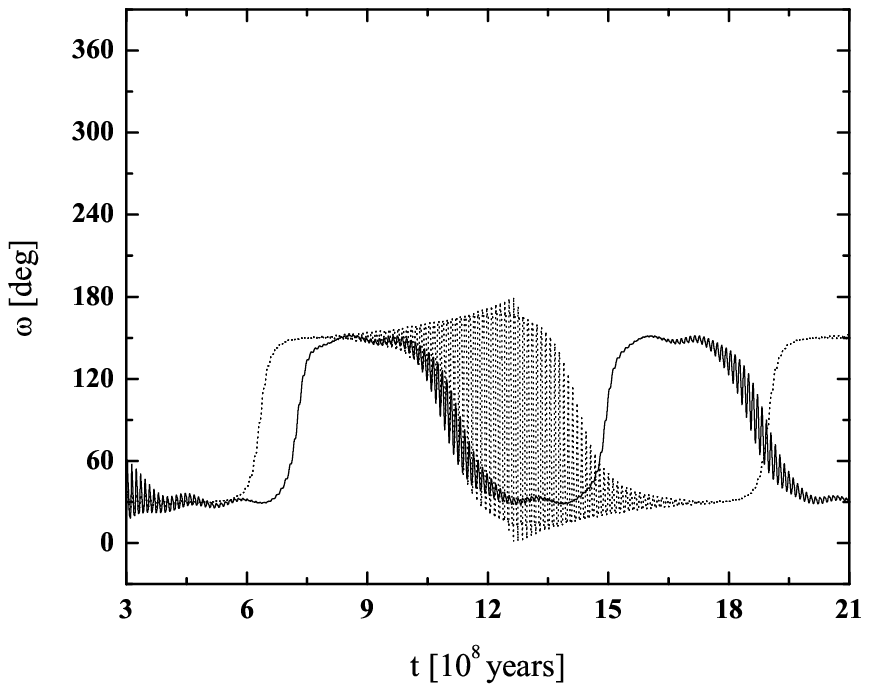} 
\label{F3}
\caption{Orbital evolution of  the comet situated in the Oort cloud with $a_{in}$ = $5 \times 10^{4}$ AU, $e_{in} \approx$ 0, $i_{in}$ = $70^{\circ}$
under the influence of the solar gravity and the galactic tide for the simple (dotted line) and the standard (solid line) models. Local density $\rho$ = 0.1 M$_{\odot}$ pc$^{-3}$
is considered for both models.}
\end{figure}

Figs. 2 and 3 compare orbital evolution of the comet calculated from the models discussed above. 

Fig. 2 shows the difference between the simple and the standard model for the the comet with initial values $a_{in}$ = $5 \times 10^{4}$ AU, 
$e_{in} \approx$ 0 and $i_{in}$ = $90^{\circ}$. For the standard model (solid line) the number of returns of the comet 
to the inner part of the Solar System (perihelion distance less than $150$ AU, approximately) is 5,
while for the simple model (dotted line) the number of returns is only 4 during the integration time 
$4.5\times 10^{9}$ years. The difference between the two current models are also shown in the evolution of the inclination $i$,
longitude of the ascending node $\Omega$ and longitude of pericenter $\omega$ in Fig. 2. 

Fig. 3 represents the case when the initial inclination of the comet is $i_{in}$ = $70^{\circ}$. 
Again, semi-major axis and eccentricity are $a_{in}$ = $5 \times 10^{4}$ AU, $e_{in} \approx$ 0.
In this special case $<di/dt>$ = 0 for the simple model (see Eq. 3). 
For the standard model the orbit of the comet can be prograde, but also retrograde.
The orbit of the comet is always prograde for $i_{in} < 90^{\circ}$. The orbit of the comet is always retrograde for $i_{in} > 90^{\circ}$.

The number of returns of the comet to the inner part of the Solar System is 6, if $i_{in}$ = $90^{\circ}$, both for the simple and standard models. 

For the difference between the number of cometary returns for the current models and more physical models compare 
Figs. 2, 3 with Figs. 5-8. 

\section{Improved models}

Current models discussed in Sec. 3 must be treated as a rough approximation to reality. At present, an improved and more realistic
physical access to galactic tide is in disposal (Kla\v{c}ka 2009a, 2009b). This section presents orbital evolution
for the improved access to galactic tide. The results are compared to the evolution for the standard model.

\subsection{Model I}

We are interested in the motion of the comet with respect to the
Sun. The Sun is moving in a distance $R_{0}$ $=$ 8 kpc from the center of the
Galaxy. Currently, the Sun is situated 30 pc above the galactic
equatorial plane ($Z_{0}$ $=$ 30 pc). Besides rotational motion
with the speed ($A$ $-$ $B$) $R_{0}$ the Sun moves with the speed
7.3 km/s in the direction normal to the galactic plane. Positional
vector of the comet with respect to the Sun is $\vec{r}$ $=$ ($\xi$,
$\eta$, $\zeta$). 

Equation of motion is taken in the form (Kla\v{c}ka 2009, Eq. 29)
\begin{eqnarray}\label{6}
\frac{d^{2} \xi}{dt^{2}} &=& - ~ \frac{G M_{\odot}}{r^{3}} ~ \xi
~+~ ( A - B ) \left [ A + B + 2 A \cos \left ( 2 ~ \omega_{0} t \right )
 \right ] ~ \xi
\nonumber \\
& & -~ 2 A ( A - B ) \sin \left ( 2 ~ \omega_{0} t \right ) ~\eta
\nonumber \\
& & +~ ( A - B )^{2} \left ( \Gamma_{1}/(b^{2}+Z_{0}^{2})^{1/2} + \Gamma_{2} 
\right ) ~R_{0} ~Z_{0}~ \cos \left ( \omega_{0} t \right)~ \zeta ~,
\nonumber \\
\frac{d^{2} \eta}{dt^{2}} &=& - ~ \frac{G M_{\odot}}{r^{3}} ~ \eta
~-~  2 A ( A - B ) \sin \left ( 2 ~ \omega_{0} t \right ) ~ \xi
\nonumber \\
& & +~ ( A - B ) \left [ A + B - 2 A \cos \left ( 2 ~ \omega_{0} t \right )
 \right ] ~ \eta
 \nonumber \\
& & -~ ( A - B )^{2} \left ( \Gamma_{1}/(b^{2}+Z_{0}^{2})^{1/2} + \Gamma_{2} 
\right ) ~R_{0} ~Z_{0}~ ~\sin \left ( \omega_{0} t \right )~ \zeta~,
\nonumber \\
\frac{d^{2} \zeta}{dt^{2}} &=& - ~ \frac{G M_{\odot}}{r^{3}} ~ \zeta
~-~ \left [ 4 ~\pi ~G ~\varrho ~+~
2 \left ( A^{2} ~-~ B^{2} \right ) \right ] ~\zeta
\nonumber \\
& & -~ 4 ~\pi ~G ~\varrho' ~
    Z_{0} \left [ \cos \left ( \omega_{0} t \right ) ~ \xi ~-~
    \sin \left ( \omega_{0} t \right ) ~ \eta \right ] ~,
\nonumber \\
\frac{d^{2} Z_{0}}{dt^{2}} &=& -~ \left [ 4 ~\pi ~G ~\varrho ~+~
2 \left ( A^{2} ~-~ B^{2} \right ) \right ] ~Z_{0} ~,
\nonumber \\
r &=& \sqrt{\xi ^{2} ~+~ \eta ^{2} ~+~ \zeta ^{2}} ~,
\nonumber \\
\omega_{0} &=& A ~-~ B ~,
\end{eqnarray}
where $G$ is the gravitational constant, $M_{\odot}$ is the mass
of the Sun and the numerical values of the other relevant quantities are
\begin{eqnarray}\label{7}
A &=& 14.25 ~\mbox{km} ~\mbox{s}^{-1} ~\mbox{kpc}^{-1} ~,
\nonumber \\
B &=& -~ 13.89 ~ \mbox{km} ~\mbox{s}^{-1} ~\mbox{kpc}^{-1} ~,
\nonumber \\
\Gamma_{1} &=& 0.084 ~\mbox{kpc}^{-1} ~,
\nonumber \\
\Gamma_{2} &=& 0.008 ~\mbox{kpc}^{-2} ~,
\nonumber \\
 \varrho &=& 0.143 ~\mbox{M}_{\odot} ~\mbox{pc}^{-3} ~,
\nonumber \\
\varrho' &=& -~ 0.0425 ~\mbox{M}_{\odot} ~\mbox{pc}^{-3} ~\mbox{kpc}^{-1} ~,
\nonumber \\
b &=& 0.25 ~\mbox{kpc} ~.
\end{eqnarray}

Eqs. (6) - (7) correspond to the model of the Galaxy presented by Dauphole et al. (1996).

\begin{figure}[h]
\centering
\includegraphics[scale=0.52]{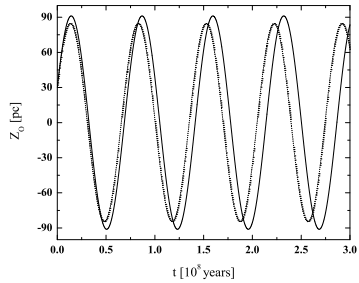}
\caption{Position of the Sun with respect to the plane of the galactic equator as a function of time. 
Galactic tides for the Model I (dotted line) and the Model II (solid line) are considered.}
\label{F4}
\end{figure}

Fig. 4 shows the position of the Sun with respect to the plane of the galactic equator as a function of time for 
two improved models. The period of oscillations of the Sun with respect to the plane of galactic 
equator is $6.96 \times 10^{7}$ years for the Model I and $7.26 \times 10^{7}$ years for the Model II. Maximal
distance of the Sun is $87.8$ pc from the plane of the galactic equator for the Model I and $91.14$ pc 
for the Model II.

The Sun is always located in the plane of galactic equator for the current models discussed in Sec. 3.

\begin{figure}[h]
\centering
\includegraphics[scale=0.52]{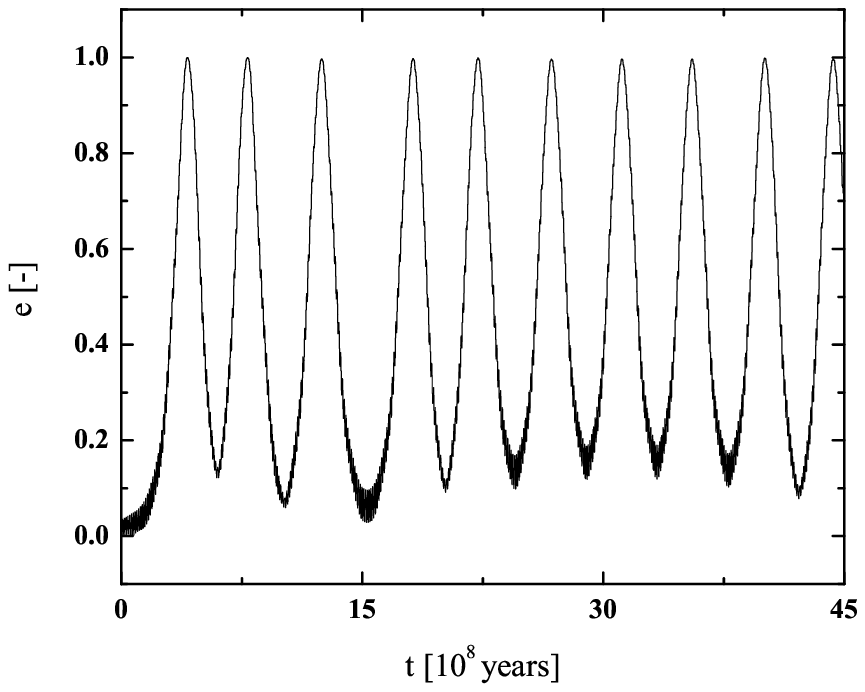}
\includegraphics[scale=0.52]{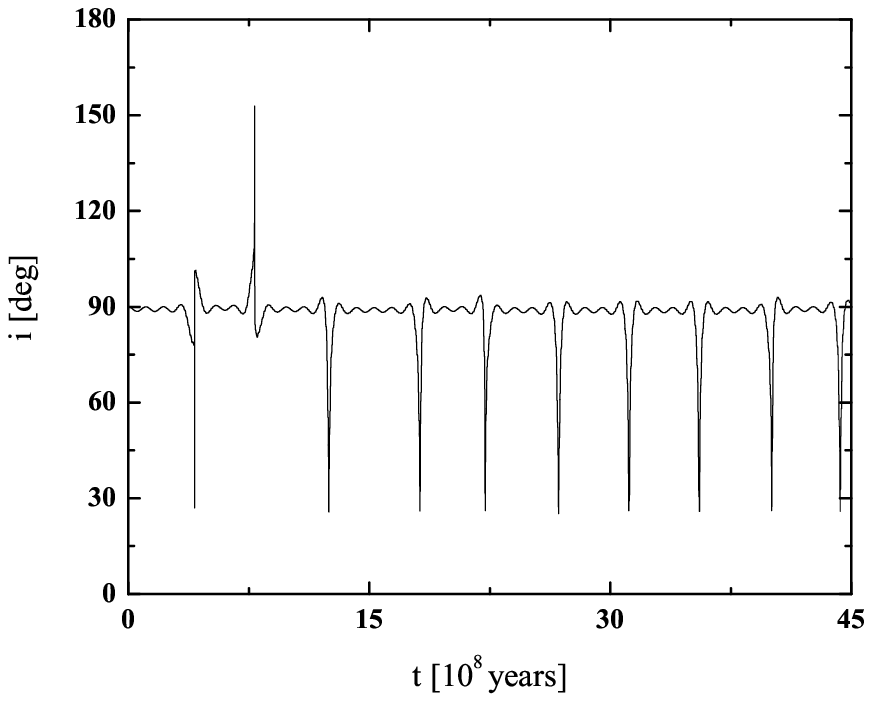}
\includegraphics[scale=0.52]{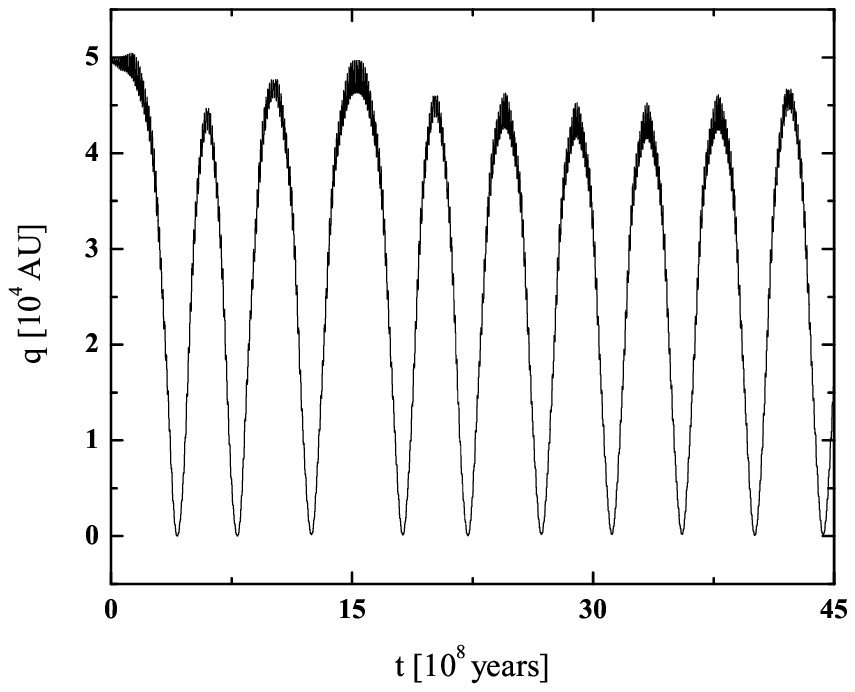}
\includegraphics[scale=0.52]{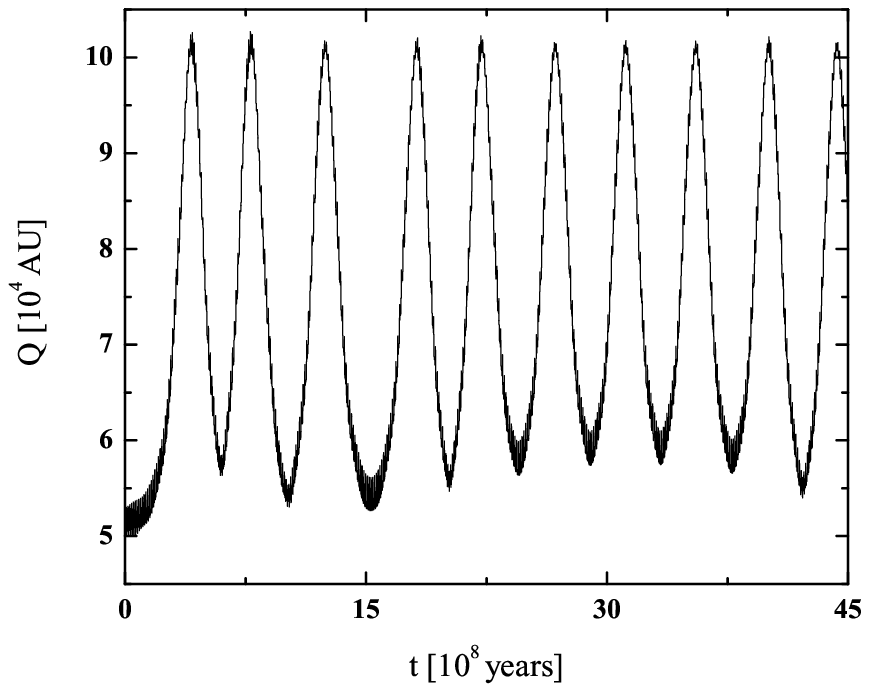}
\includegraphics[scale=0.52]{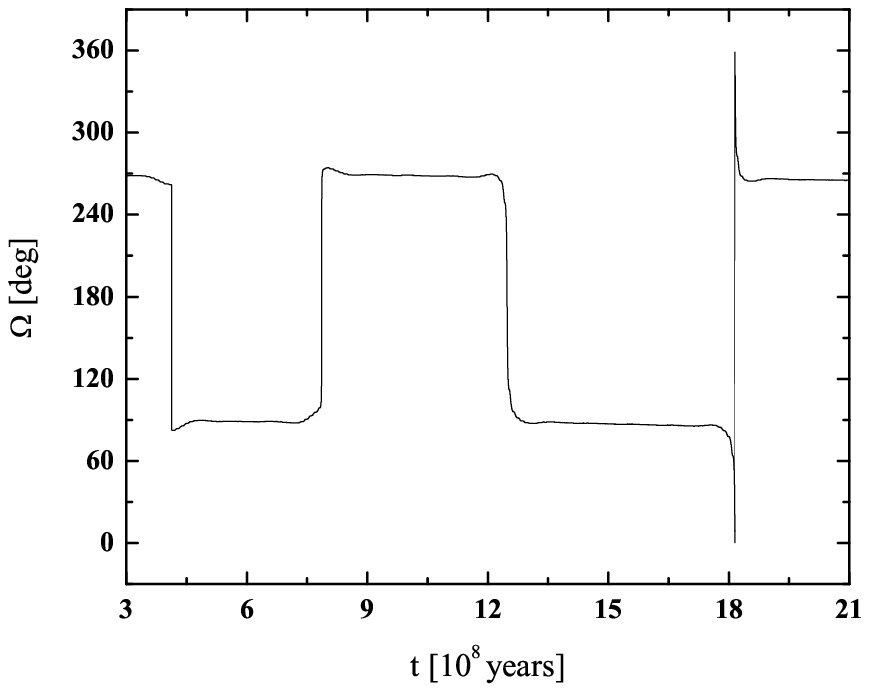}
\includegraphics[scale=0.52]{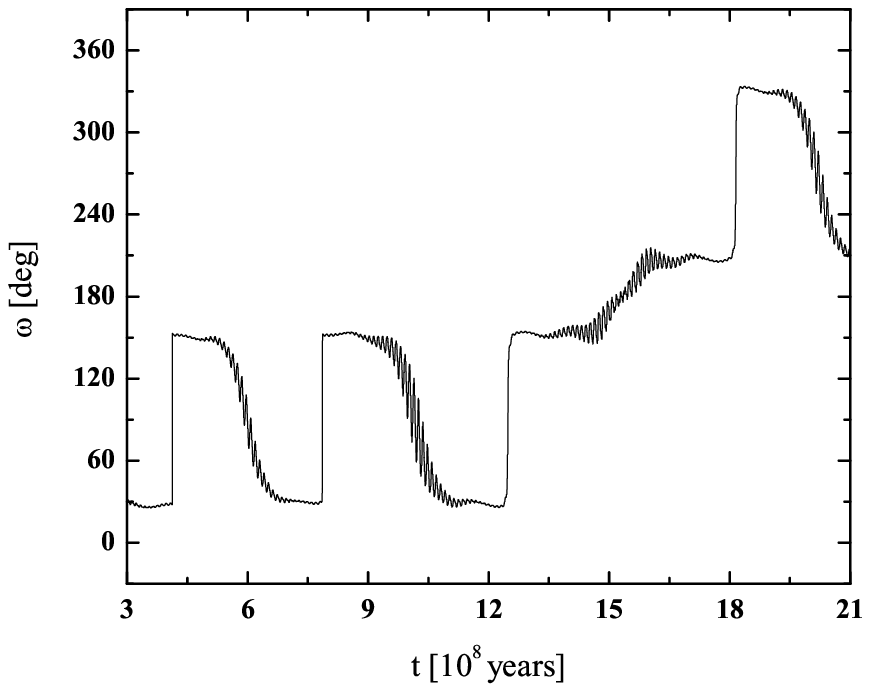}
\label{F5}
\caption{Orbital evolution of the comet situated in the Oort cloud with $a_{in}$ = $5 \times 10^{4}$ AU, $e_{in} \approx 0$, $i_{in}$ = $90^{\circ}$
under the influence of the solar gravity and the galactic tide for the Model I.}
\end{figure}
\begin{figure}[h]
\centering
\includegraphics[scale=0.52]{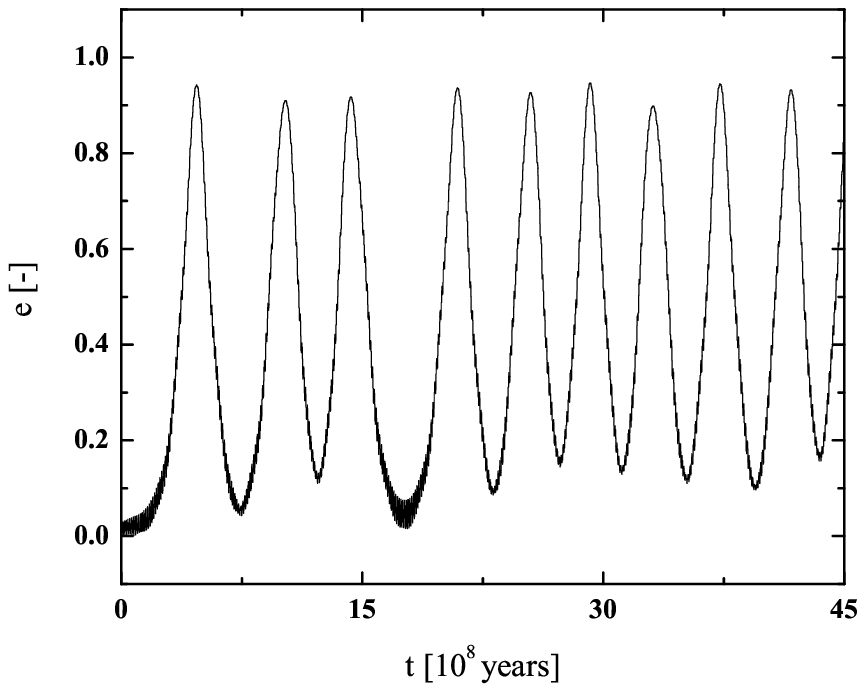}
\includegraphics[scale=0.52]{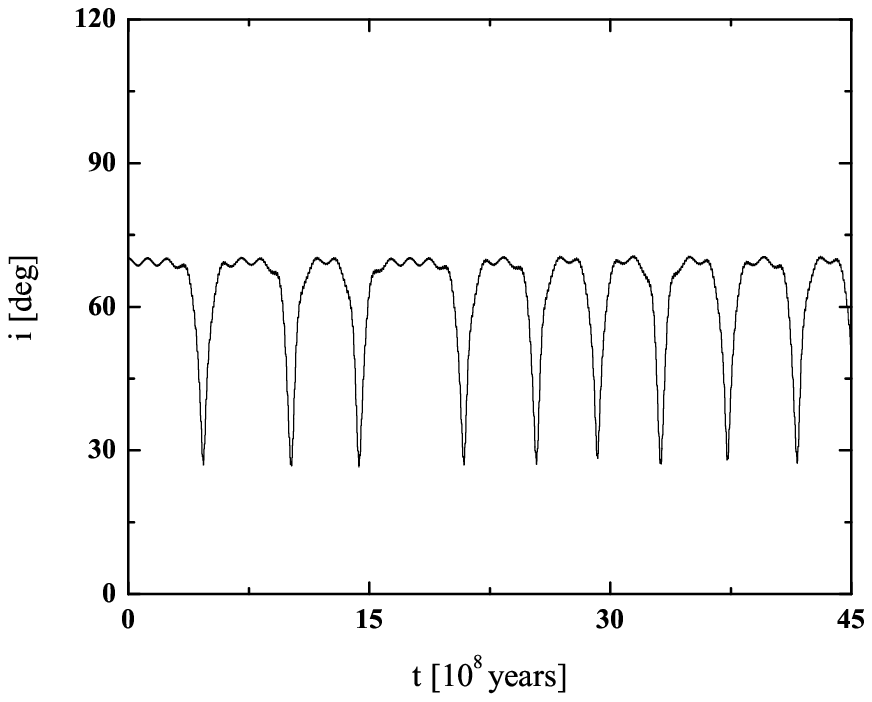}
\includegraphics[scale=0.52]{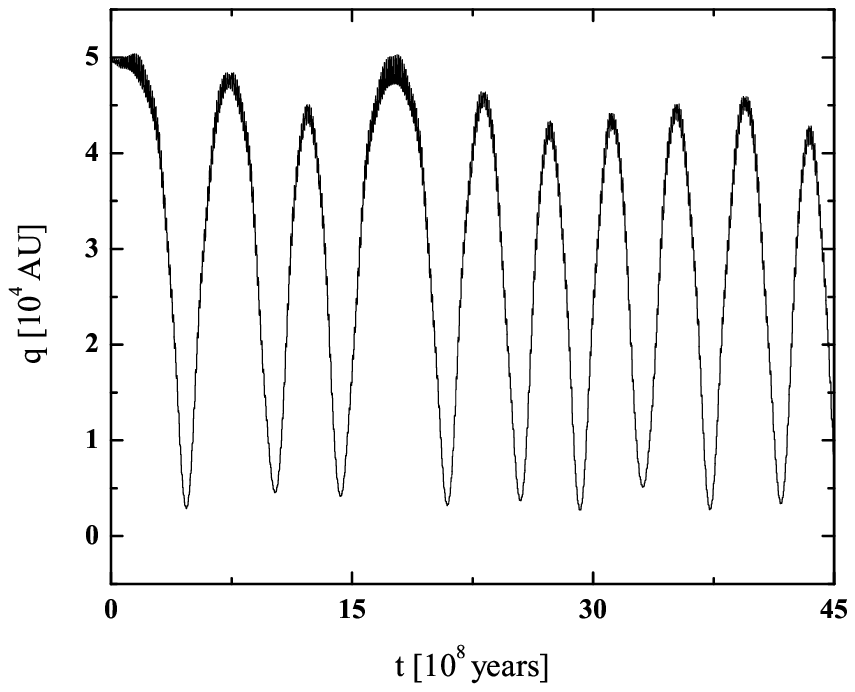}
\includegraphics[scale=0.52]{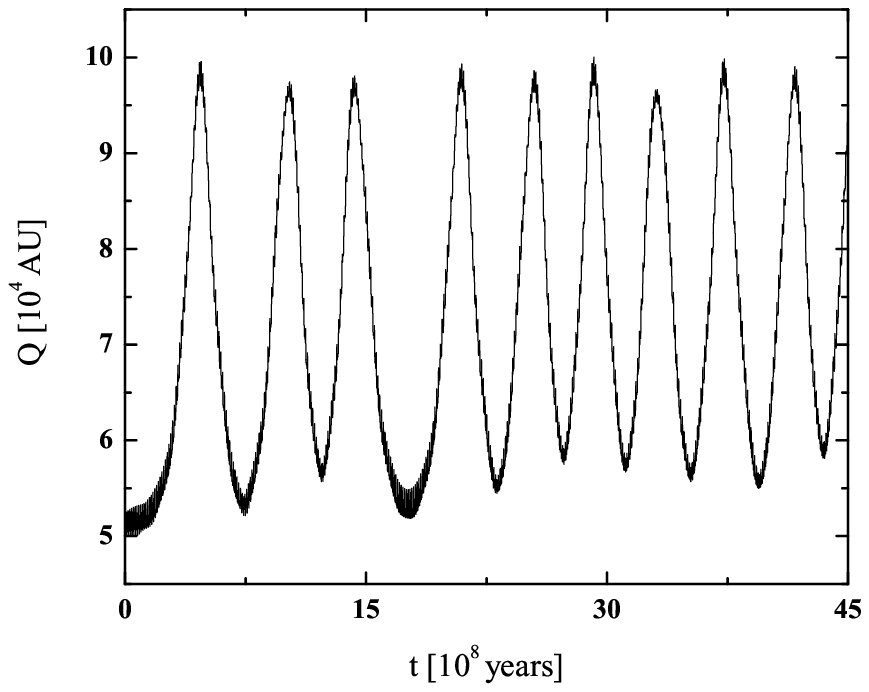}
\includegraphics[scale=0.52]{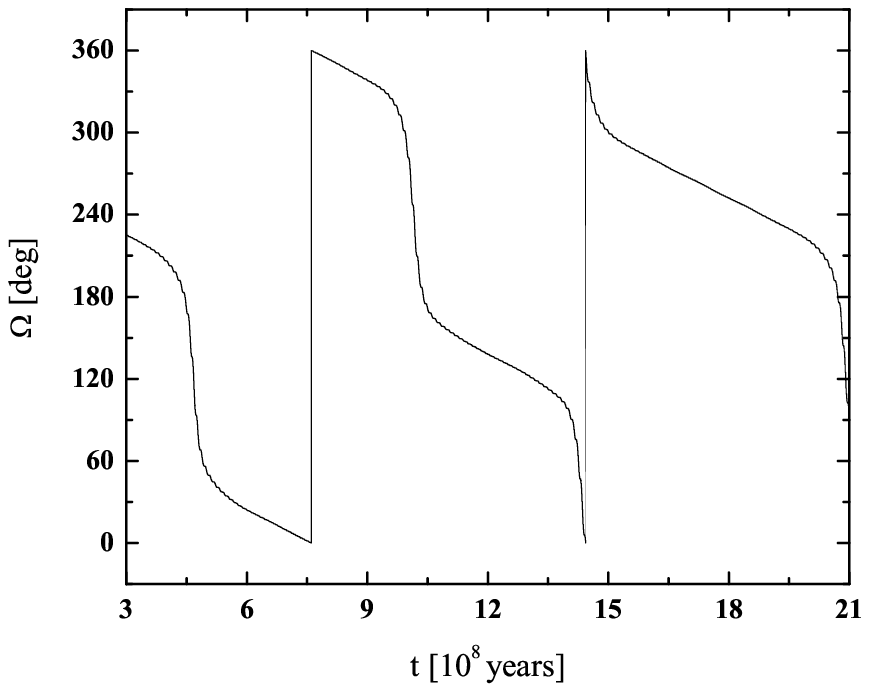}
\includegraphics[scale=0.52]{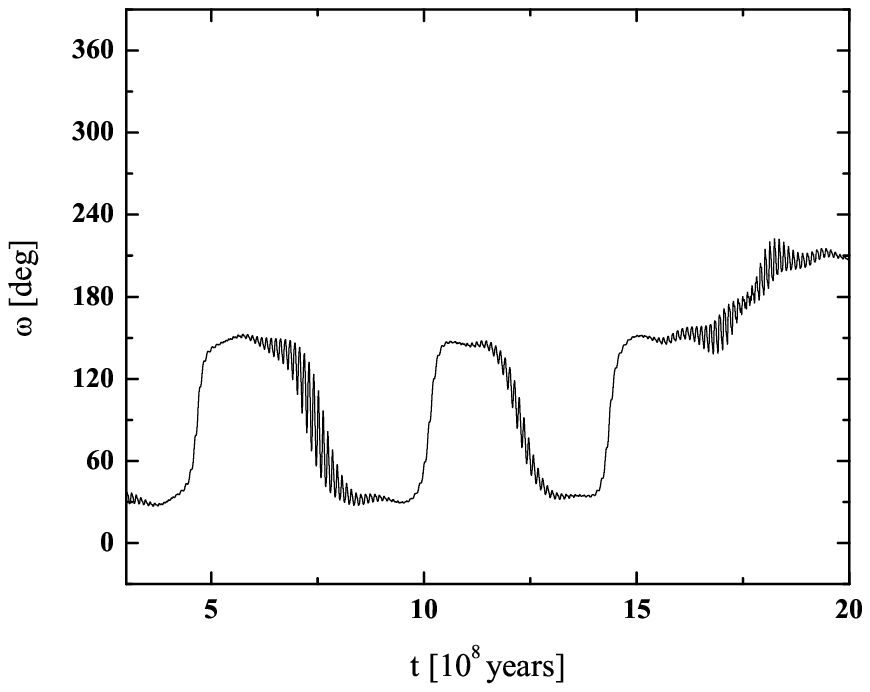}
\label{F6}
\caption{Orbital evolution of the comet situated in the Oort cloud with $a_{in}$ = $5 \times 10^{4}$ AU, $e_{in} \approx$ 0, $i_{in}$ = $70^{\circ}$
under the influence of the solar gravity and the galactic tide for the Model I.}
\end{figure}

Figs. 5 and 6 depict the orbital evolution of the comet as a numerical solution of Eq. (6), if also Eqs. (1) and (7) are taken into account.
Comet is initially situated on the almost circular orbit ($e_{in} \approx$ 0) in the Oort cloud, $a_{in}$ = $5 \times 10^{4}$ AU. 
Orbital evolutions for two different initial inclinations are presented, $i_{in}$ = $90^{\circ}$ in Fig. 5 and $i_{in}$ = $70^{\circ}$ in Fig. 6.
For comparison, the initial conditions of the comet are the same as for numerical calculations presented in Sec. 3.

Figs. 5 and 6 represent the evolution of eccentricity $e$, inclination $i$, , longitude of the ascending node $\Omega$, longitude of pericenter $\omega$,
perihelion distance $q$ and aphelion distance $Q$ during 4.5 $\times$ 10$^{9}$ years. 
Semimajor axes are practically constant during the whole integration time for both presented inclinations.

Evolutions of the orbital elements in Figs. 5, 6 show the main differences between the current models described in Sec 3 
and our improved Model I (compare with Figs. 2, 3 ).  

\begin{table}[h]
\centering
\begin{tabular}{|c|c|c|c|c|}
\hline
 & \multicolumn{2}{|c|} {Standard model} & \multicolumn {2} {c|} {Model I} \\
\hline
$i_{in}$ &  $<q>$ [AU] & $<Q>$ [AU] & $<q>$ [AU] & $<Q>$ [AU] \\
\hline
$70^{\circ}$ & 33792 & 66889 & 29484 & 71526  \\
\hline
$90^{\circ}$ & 29827 & 70685 & 27574 & 73605 \\
\hline
$120^{\circ}$ & 36614 & 63922 & 32934 & 67884  \\
\hline
\end{tabular}
\caption{Mean values of the perihelion and aphelion distances during the integration time $4.5\times 10^{9}$ years 
for the standard model and the Model I. Results for three different initial inclinations of the comet 
with $a_{in}$ = $5 \times 10^{4}$ AU, $e_{in} \approx$ 0 are presented.}
\label{tab:1}
\end{table}

Table 1 presents mean values of the perihelion and aphelion distances for the standard model and the Model I.
Various initial inclinations are presented. Both the standard model and the Model I have minima of the mean values 
of the perihelion distances for the initial inclination $i_{in}$ = $90^{\circ}$. Also the maxima of the mean values of
the aphelion distances belong to the initial inclination $i_{in}$ = $90^{\circ}$. 
On the basis of our calculations we can state that the Model I shows lower mean values of the perihelion distances 
than the standard model and this holds irrespective to initial inclinations. 
The Model I also shows higher mean values of the aphelion distances than 
the standard model, again irrespective of the initial inclinations.

\subsection{Model II}

Equation of motion is taken in the form (Kla\v{c}ka 2009a, Eqs. 26-27):
\begin{eqnarray}\label{8}
\frac{d^{2} \xi}{dt^{2}} &=& - ~ \frac{G M_{\odot}}{r^{3}} ~ \xi
~+~ ( A - B ) \left [ A + B + 2 A \cos \left ( 2 ~ \omega_{0} t \right )
 \right ] ~ \xi
\nonumber \\
& & -~ 2 A ( A - B ) \sin \left ( 2 ~ \omega_{0} t \right ) ~\eta
\nonumber \\
& & +~ 2 ( A - B )^{2} \left ( \Gamma_{1} - \Gamma_{2} Z_{0} ^{2}
\right ) ~R_{0} ~Z_{0}~ \cos \left ( \omega_{0} t \right)~ \zeta ~,
\nonumber \\
\frac{d^{2} \eta}{dt^{2}} &=& - ~ \frac{G M_{\odot}}{r^{3}} ~ \eta
~-~  2 A ( A - B ) \sin \left ( 2 ~ \omega_{0} t \right ) ~ \xi
\nonumber \\
& & +~ ( A - B ) \left [ A + B - 2 A \cos \left ( 2 ~ \omega_{0} t \right )
 \right ] ~ \eta
 \nonumber \\
& & -~ 2 ( A - B )^{2} \left ( \Gamma_{1} - \Gamma_{2} Z_{0} ^{2}
\right ) ~R_{0} ~Z_{0}~ ~\sin \left ( \omega_{0} t \right )~ \zeta~,
\nonumber \\
\frac{d^{2} \zeta}{dt^{2}} &=& - ~ \frac{G M_{\odot}}{r^{3}} ~ \zeta
~-~ \left [ 4 ~\pi ~G ~\varrho ~+~
2 \left ( A^{2} ~-~ B^{2} \right ) \right ] ~\zeta
\nonumber \\
& & -~ 4 ~\pi ~G ~\varrho' ~
    Z_{0} \left [ \cos \left ( \omega_{0} t \right ) ~ \xi ~-~
    \sin \left ( \omega_{0} t \right ) ~ \eta \right ] ~,
\nonumber \\
\frac{d^{2} Z_{0}}{dt^{2}} &=& -~ \left [ 4 ~\pi ~G ~\varrho ~+~
2 \left ( A^{2} ~-~ B^{2} \right ) \right ] ~Z_{0} ~,
\nonumber \\
r &=& \sqrt{\xi ^{2} ~+~ \eta ^{2} ~+~ \zeta ^{2}} ~,
\nonumber \\
\omega_{0} &=& A ~-~ B ~,
\end{eqnarray}
where $G$ is the gravitational constant, $M_{\odot}$ is the mass
of the Sun and the numerical values of the other relevant quantities are
\begin{eqnarray}\label{9}
A &=& 14.2 ~\mbox{km} ~\mbox{s}^{-1} ~\mbox{kpc}^{-1} ~,
\nonumber \\
B &=& -~ 12.4 ~\mbox{km} ~\mbox{s}^{-1} ~\mbox{kpc}^{-1} ~,
\nonumber \\
\Gamma_{1} &=& 0.124 ~\mbox{kpc}^{-2} ~,
\nonumber \\
\Gamma_{2} &=& 1.586 ~\mbox{kpc}^{-4} ~,
\nonumber \\
 \varrho &=& 0.130 ~\mbox{M}_{\odot} ~\mbox{pc}^{-3} ~,
\nonumber \\
\varrho' &=& -~ 0.037 ~\mbox{M}_{\odot} ~\mbox{pc}^{-3} ~\mbox{kpc}^{-1} ~,
\end{eqnarray}
see Eqs. (20)-(21) in Kla\v{c}ka (2009). If one wants to use other values
of the Oort constants $A$ and $B$, then he can use the following equation
for calculation of mass density in the neighborhood of the Sun:
\begin{eqnarray}\label{10}
\varrho &=&  \varrho_{disk} ~+~ \varrho_{halo} ~,
\nonumber \\
\varrho_{disk} &=& 0.126 ~\mbox{M}_{\odot} ~\mbox{pc}^{-3} ~,
\nonumber \\
\varrho_{halo} &=& (4 \pi G)^{-1} [ X(Galaxy) - X(disk) - X(bulge) ]  ~,
\nonumber \\
X(Galaxy) &\equiv& - (A - B) \times (A + 3 B)
\nonumber \\
X(disk) &=&  396.898 ~\mbox{km}^{2} ~\mbox{s}^{-2} ~\mbox{kpc}^{-2} ~,
\nonumber \\
X(bulge) &=& 0.625 ~\mbox{km}^{2} ~\mbox{s}^{-2} ~\mbox{kpc}^{-2} ~.
\end{eqnarray}
The value $X(Galaxy)$ $=$ 611.800 km$^{2}$ s$^{-2}$ kpc$^{-2}$ holds for
$A$ $=$ 14.2 km s$^{-1}$ kpc$^{-1}$ and
$B$ $=$ $-$ 12.4 km s$^{-1}$ kpc$^{-1}$.
Moreover, Eq. (17) of Kla\v{c}ka (2009) can be used. \\

\begin{figure}[h]
\centering
\includegraphics[scale=0.52]{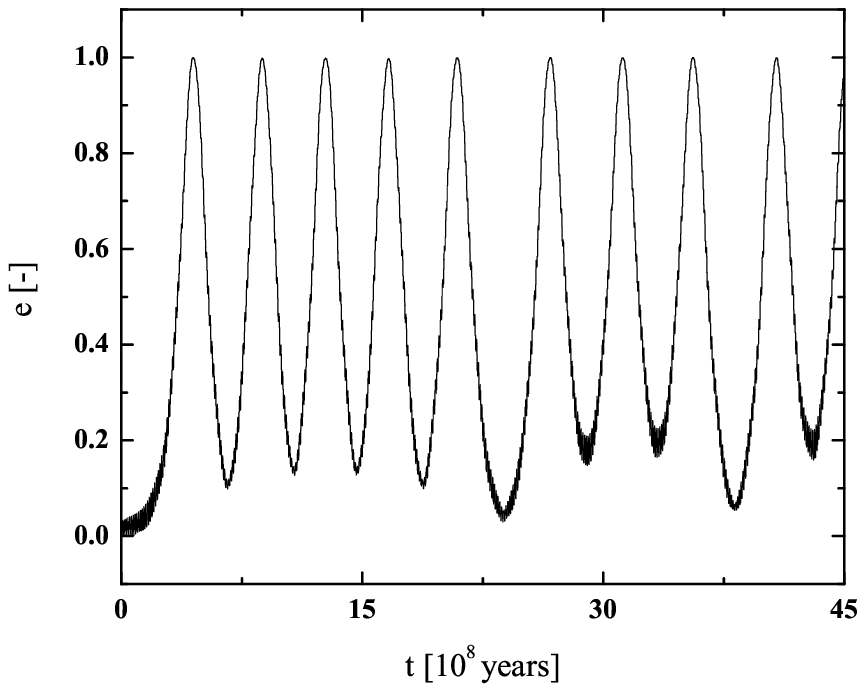}
\includegraphics[scale=0.52]{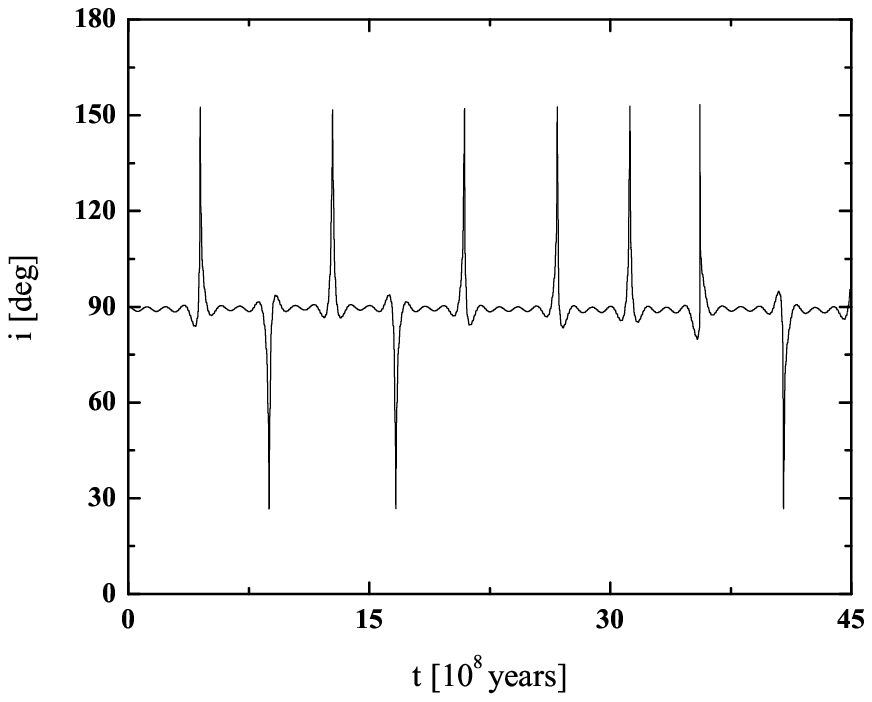}
\includegraphics[scale=0.52]{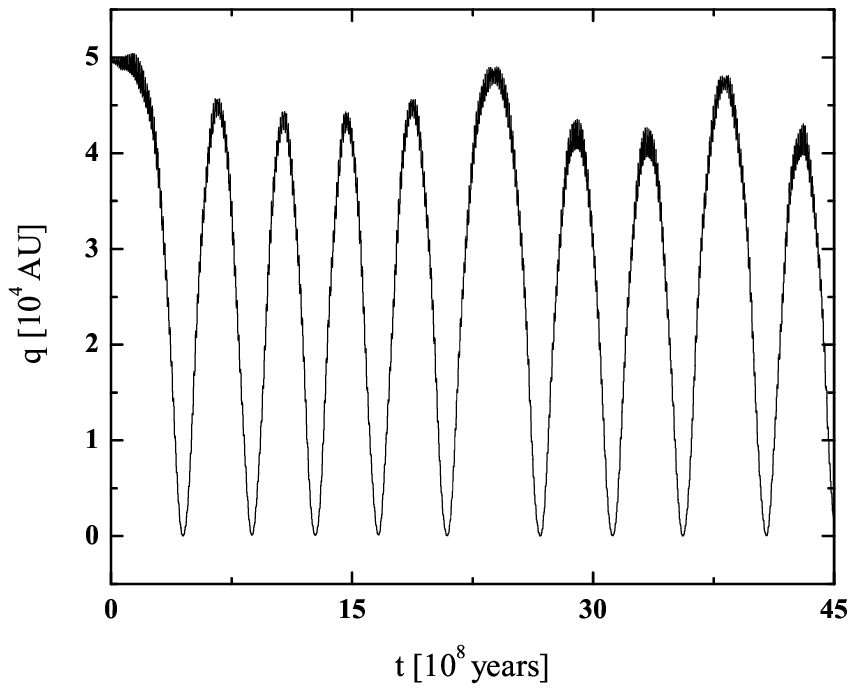}
\includegraphics[scale=0.52]{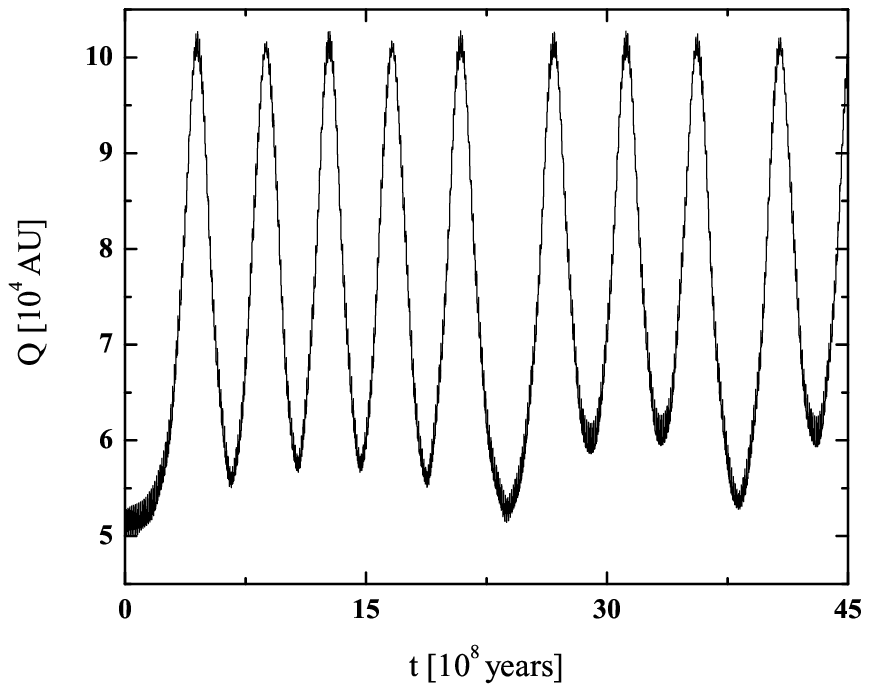}
\includegraphics[scale=0.52]{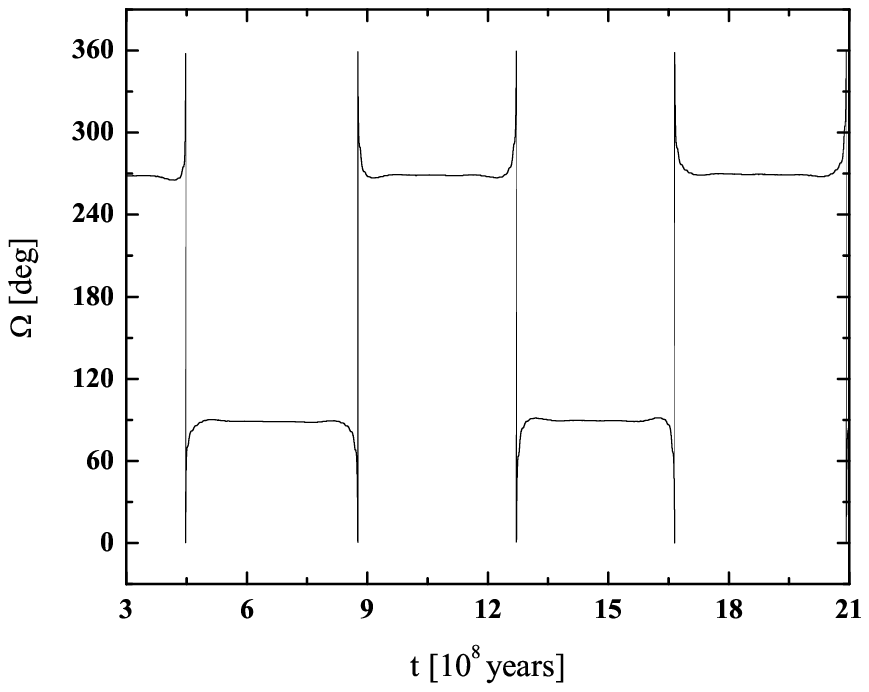}
\includegraphics[scale=0.52]{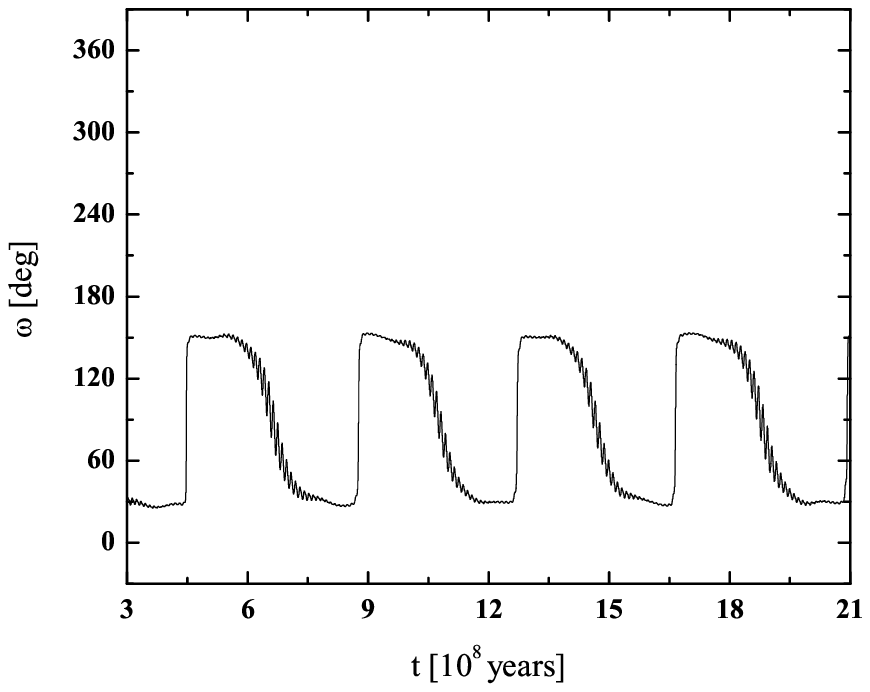}
\label{F7}
\caption{Orbital evolution of the comet with initial values $a_{in}$ = $5 \times 10^{4}$ AU, $e_{in} \approx$ 0, $i_{in}$ = $90^{\circ}$
under the influence of the solar gravity and the galactic tide for the Model II}
\end{figure}

\begin{figure}[h]
\centering
\includegraphics[scale=0.52]{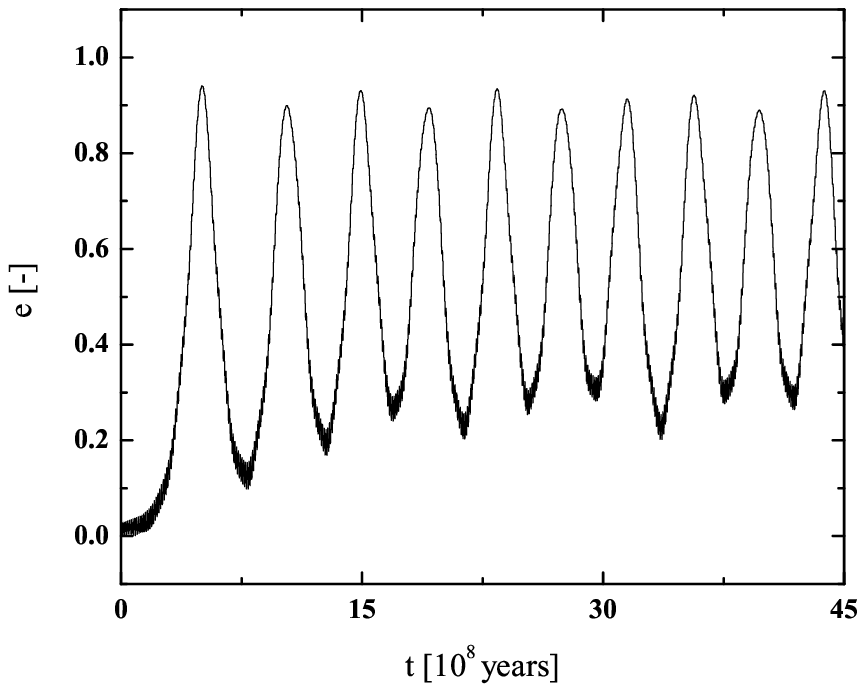}
\includegraphics[scale=0.52]{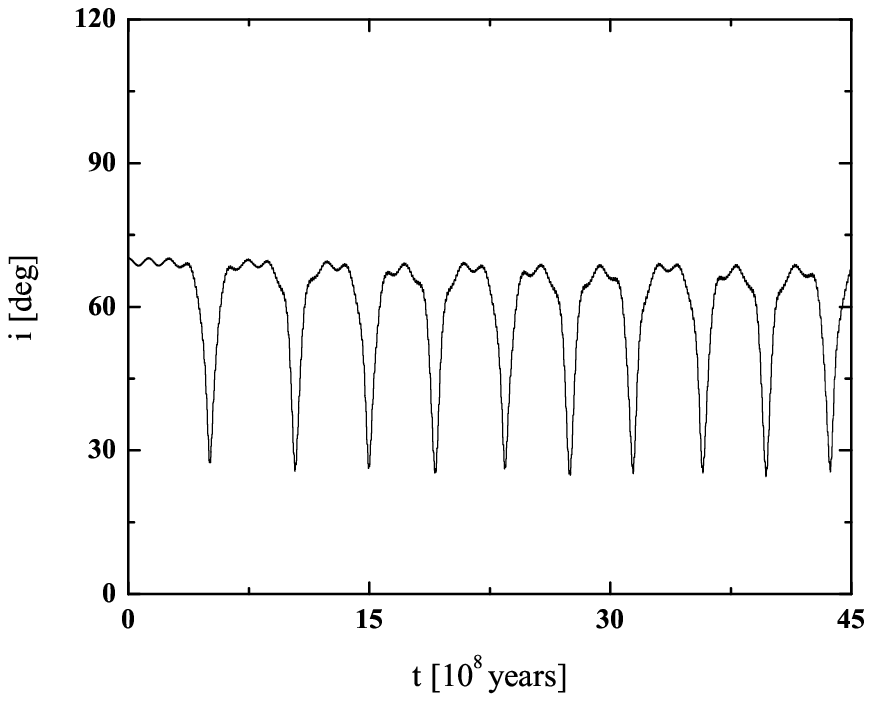}
\includegraphics[scale=0.52]{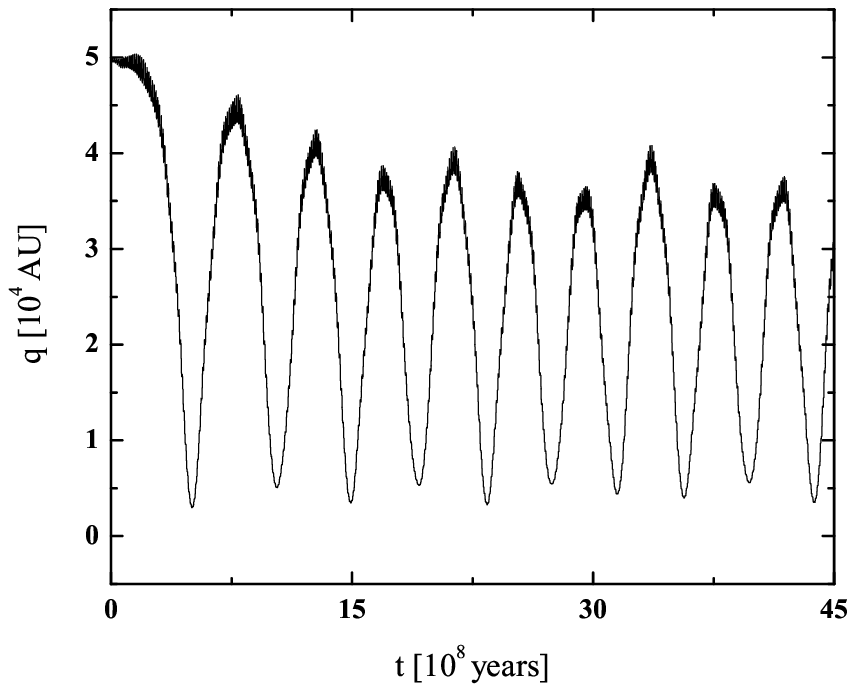}
\includegraphics[scale=0.52]{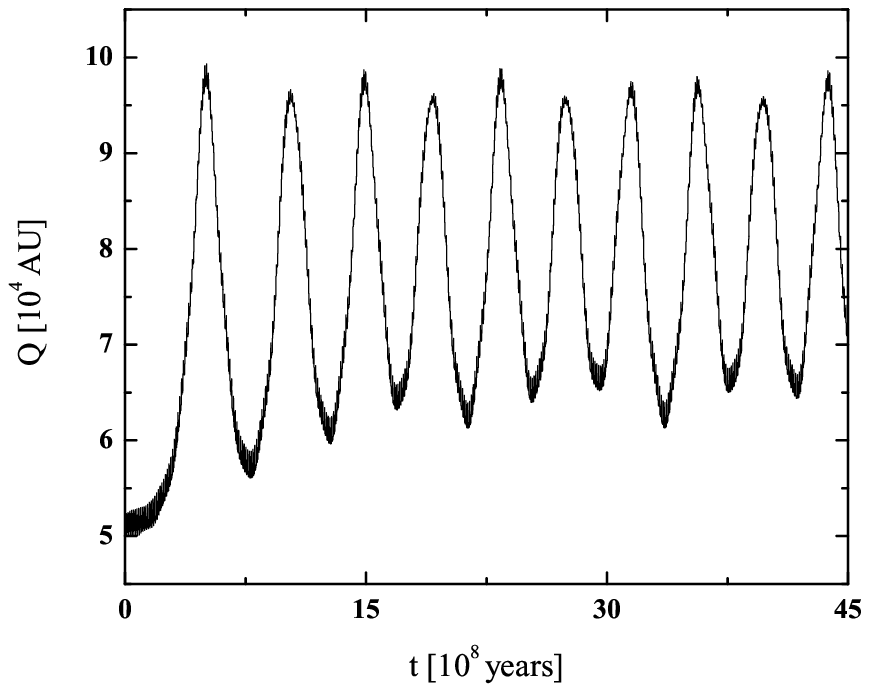}
\includegraphics[scale=0.52]{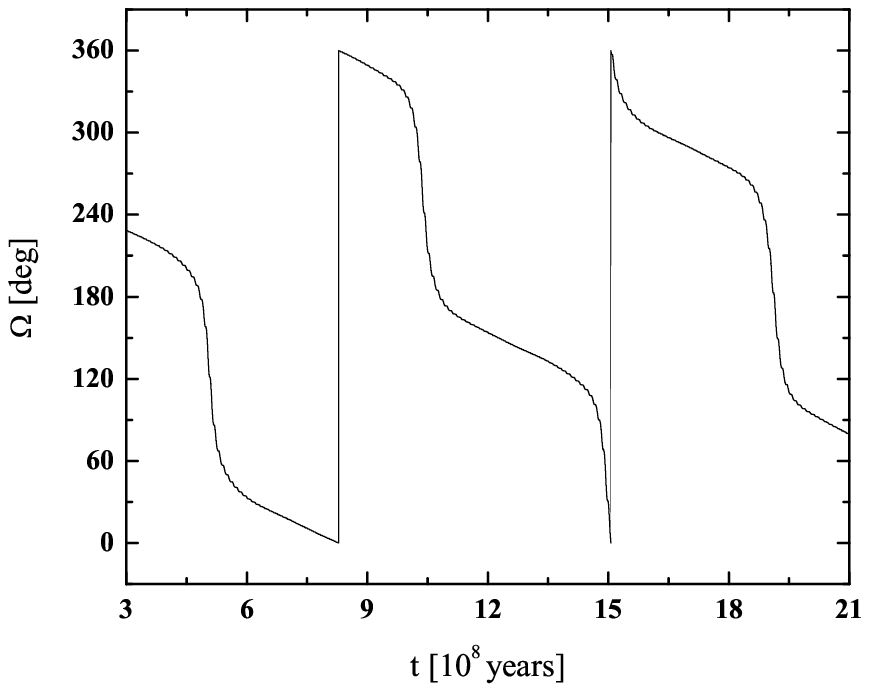}
\includegraphics[scale=0.52]{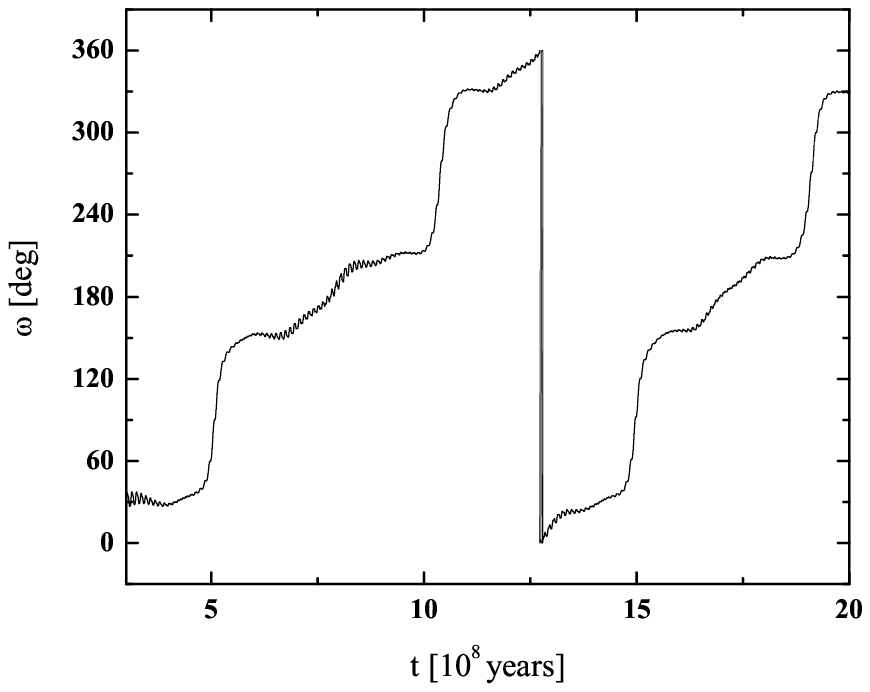}
\label{F8}
\caption{Orbital evolution of the comet with $a_{in}$ = $5 \times 10^{4}$ AU, $e_{in} \approx$ 0, $i_{in}$ = $70^{\circ}$
under the influence of the solar gravity and the galactic tide for the Model II.}
\end{figure} 

Figs. 7 and 8 represent the orbital evolution of a comet as a numerical solution of Eq. (8), if also Eqs. (1) and (9) are taken into account.
The comet is initially situated in the Oort cloud ( $a_{in}$ = $5 \times 10^{4}$ AU) on the initially almost circular orbit as well as in pervious
cases. Orbital evolutions for two different initial inclinations are presented, $i_{in}$ = $90^{\circ}$ in Fig. 7, $i_{in}$ = $70^{\circ}$ in Fig. 8.

Figs. 7 and 8 depict the evolution of eccentricity $e$, inclination $i$, , longitude of the ascending node $\Omega$, longitude of pericenter $\omega$,
perihelion distance $q$ and aphelion distance $Q$ during the current lifetime of the Solar System. 
Semimajor axes are practically constant during the whole integration time for both presented inclinations.

The evolutions of the orbital elements in Figs. 7 and 8 show the main differences between the current models described in Sec. 3 
and our improved physical Model II (compare with Figs. 2 and 3 ). \\

\begin{table}[h]
\centering
\begin{tabular}{|c|c|c|c|c|}
\hline
 & \multicolumn{2}{|c|} {Standard model} & \multicolumn {2} {c|} {Model II} \\
\hline
$i_{in}$ &  $<q>$ [AU] & $<Q>$ [AU] & $<q>$ [AU] & $<Q>$ [AU] \\
\hline
$70^{\circ}$ & 33792 & 66889 & 25157 & 75766  \\
\hline
$90^{\circ}$ & 29827 & 70685 & 27306 & 73757  \\
\hline
$120^{\circ}$ & 36614 & 63922 & 33768 & 66971 \\
\hline
\end{tabular}
\caption{Mean values of the perihelion and aphelion distances during the integration time $4.5\times 10^{9}$ years 
for the standard model and the Model II. Three different initial inclinations of the comet 
with $a_{in}$ = $5 \times 10^{4}$ AU, $e_{in} \approx$ 0 are considered.}
\label{tab:2}
\end{table}

Table 2 contains mean values of the perihelion and aphelion distances for the standard model and the Model II.
Various initial inclinations are presented as well as in Table 1.
For the standard model, the minimum of the mean value of the perihelion distance belongs to the initial inclination 
$i_{in}$ = $90^{\circ}$, but for the Model II the least value belongs to $i_{in}$ = $70^{\circ}$. 
Maximum of the mean value of the aphelion distance belongs to the initial inclination $i_{in}$ = $90^{\circ}$ 
for the standard model and to $i_{in}$ = $70^{\circ}$ for the Model II. The Model II shows lower 
mean values of the perihelion distances than the standard model for all of the presented initial inclinations. 
The Model II also shows higher mean values of the aphelion distances than the standard model for all 
of the presented initial inclinations.

\section{Discussion} 
Models I and II are based on more realistic physics than the standard or simple models.
The Model I corresponds to the model of the Galaxy by Dauphole et al. (1986).
This model of the Galaxy does not respect several important observational facts.
It does not yield a flat rotation curve for great galactocentric distances.
Mass density in the vicinity of the Sun is relatively large in comparison to
newer values. Moreover, it does not produce the observed values of the Oort constants.
Both of these observational facts are taken into account in the model of the Galaxy which
served for our Model II (Kla\v{c}ka 2009a). This is the reason why we consider 
the Model II to be more relevant than the Model I. In any case, both of these
models take into account motion of the Sun in a more realistic way than it is considered
in the simple and standard models. 

As a consequence of the previous discussion we can stress several facts. The standard model
yields mean value of the perihelion distance in  $\approx$ 35$\%$ higher than the real 
value for $i_{in}$ $=$ 70$^{\circ}$. This follows from  Tables 1 and 2: 
$< q >$ (standard model) / $<q>$ (Model I) = 1.146,
$< q >$ (standard model) / $<q>$ (Model II) = 1.343.
Moreover, the Model II is the only of the discussed models which yields
$\langle q \rangle (i_{in} = 70^{\circ})$ $<$ 
$\langle q \rangle (i_{in} = 90^{\circ})$.

Let $\tau_{q}$ is a timescale on which a comet's perihelion changes
in $\Delta q$. Let us consider Fig. 7 and Table 3 for finding $\tau_{q}$ for
$t \approx$ 2.3 $\times$ 10$^{9}$ years. The case of maximum in
$q = q(t)$ yields: \\
$\tau_{q}$ ($q =$ 4.8 $\times$ 10$^{4}$ AU, $\Delta q$ $=$ 0.4 $\times$
10$^{4}$ AU; reality) $=$ 9.4 $\times$ 10$^{7}$ yrs, \\
while the relation presented by Levison and Dones (2007, p. 583) yields \\
$\tau_{q}$ ($q =$ 4.8 $\times$ 10$^{4}$ AU, $\Delta q$ $=$ 0.4 $\times$
10$^{4}$ AU; literature) $=$ 4.8 $\times$ 10$^{6}$ yrs. \\
The mathematical relation yields 20-times smaller value than the real value.
Now, let us consider minimum in $q = q(t)$ for
$t \approx$ 2.3 $\times$ 10$^{9}$ years. The real value is \\
$\tau_{q}$ ($q =$ 1/4 $\times$ 10$^{4}$ AU, $\Delta q$ $= $ 1/2 $\times$
10$^{4}$ AU; reality) $=$ 4.7 $\times$ 10$^{7}$ yrs, \\
while the mathematical relation yields \\
$\tau_{q}$ ($q =$ 1/4 $\times$ 10$^{4}$ AU, $\Delta q$ $=$ 1/2 $\times$
10$^{4}$ AU; literature) $=$ 2.6 $\times$ 10$^{7}$ yrs, \\
or, using the value $q$ $=$ 38 AU from Table 3, \\
$\tau_{q}$ ($q =$ 38 AU, $\Delta q$ $=$ 1/2 $\times$
10$^{4}$ AU; literature) $=$ 2.1 $\times$ 10$^{8}$ yrs. \\
Of course, smaller value of $\Delta q$ may be taken into account.
We can say that \\
0.5 $<$ $\tau_{q}$ (literature) / $\tau_{q}$ (reality) $<$ 20.

\begin{figure}[h]
\centering
\includegraphics[scale=0.52]{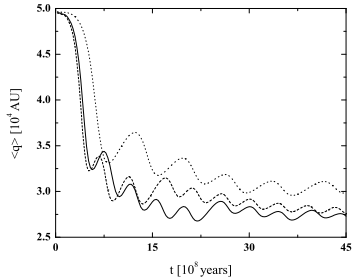}
\label{F9}
\caption{Comparison of the evolution of the mean perihelion distance of the comet with $a_{in}$ = $5 \times 10^{4}$ AU, $e_{in} \approx$ 0, $i_{in}$ = $90^{\circ}$ for the standard model (dotted line),
the Model I (dashed line) and the Model II (solid line).}
\end{figure}

\begin{table}[h]
\centering
\begin{tabular}{|r|r|r|r|r|r|}
\hline
\multicolumn {2} {|c|} {Standard model} & \multicolumn{2}{|c|} {Model I} & \multicolumn{2}{c|} {Model II} \\
\hline
q [AU] & Q [AU] & q [AU] & Q [AU] & q [AU] & Q [AU] \\
\hline
113.0 & 101142.6 & 1.5 & 102594.9 &	24.7 &	102713.9 \\
\hline
38.7 & 102137.5 & 143.8 &	102692.2 &	101.3 &	101620.5 \\
\hline
3.7 & 101971.8 & 115.7 &	101718.5 &	93.4 &	102701.1 \\
\hline
71.9 & 101056.8 & 50.5 &	102034.1 &	104.2 &	101700.1 \\
\hline
21.2 & 101258.6 & 187.5 &	102263.4 &	38.1 &	102769.8 \\ 
\hline
111.6 & 100130.6 & 194.9 &	101536.0 &	25.9 &	102648.5  \\
\hline 
-- & -- & 192.1 &	101758.7 &	43.7 &	102744.2 \\
\hline
-- & -- & 90.5 &	101707.7 &	1.0 &	102455.4 \\
\hline
-- & -- & 149.0 &	102142.4 &	37.5 &	102046.6 \\
\hline
-- & -- & --   &	101552.9 &	--	&	-- \\
\hline
\end{tabular}
\caption{Extremal perihelion and aphelion distances of the comet with 
$a_{in}$ = $5 \times 10^{4}$ AU, $e_{in} \approx 0$, $i_{in}$ = $90^{\circ}$ 
during the integration time $4.5\times 10^{9}$ years for the standard model, the Model I and the Model II.}
\label{tab:3}
\end{table}

\begin{table}[h]
\centering
\begin{tabular}{|r|r|r|r|r|r|}
\hline
\multicolumn {2} {|c|} {Standard model} & \multicolumn{2}{|c|} {Model I} & \multicolumn{2}{c|} {Model II} \\
\hline
q [AU] & Q [AU] & q [AU] & Q [AU] & q [AU] & Q [AU] \\
\hline
3465.38	& 98209.0 & 2883.6 	& 99568.6 &	2968.3 & 99308.4 \\
\hline
3036.7 & 98793.8 & 4539.5 & 97433.9 & 5067.2 & 96652.2 \\
\hline
3063.5 & 98775.0 & 4168.9 & 98050.3 & 3446.5 & 98680.8 \\
\hline
4466.9 & 96966.1 & 3190.1 & 99294.7 & 5299.0 & 96224.7 \\
\hline
3062.8 & 98648.0 & 3689.2 & 98594.1 & 3267.2 & 98839.0 \\
\hline
-- & 	-- & 	2714.9 & 100020.7 & 5441.0 & 95944.8 \\
\hline
-- &	-- & 	5114.2 & 96658.8 & 4385.5 & 97489.7 \\
\hline
-- &	-- &	2753.9 & 99851.9 & 3970.5 & 97988.4 \\
\hline
-- & 	-- &	3407.8 & 99033.9 & 5573.5 & 95894.2 \\
\hline
\end{tabular}
\caption{Extremal perihelion and aphelion distances of the comet with 
$a_{in}$ = $5 \times 10^{4}$ AU, $e_{in} \approx 0$, $i_{in}$ = $70^{\circ}$ 
during the integration time $4.5\times 10^{9}$ years for the standard model, the Model I and the Model II.}
\label{tab:4}
\end{table}

Table 3 presents minimal values of perihelion distances for the calculated models.
While the standard model yields three cases with $q_{min}$ $<$ 50 AU, the Model I 
shows that the same situation occurs in one case, only. As we have discussed above, the
relevant Model II yields six cases with $q_{min}$ $<$ 50 AU.
Table 4  presents values of the same quantities for initial galactocentric inclination 70 degrees.
As a result we can state that the smallest values of $q_{min}$ exist for initial inclination
corresponding to 90 degrees. Fig. 9 depicts time average of evolution of perihelion distance
$\langle q \rangle$ $\equiv$ $\langle q(t) \rangle$ $=$ $\int_{0}^{t} q(t') dt'$ / $t$. The evolution of perihelion distance yields minimal
values for the best physical model represented by our Model II.

The effect of Solar System's vertical motion above and below the galactic plane on the possible
terrestrial mass extinctions was discussed in the past (see, e.g., Weissman 1990, 
Rampino and Stothers 1984, Schwartz and James 1984). Results corresponding to Fig. 4
show that period of the Solar System's vertical oscillations is 72.6 $\times$ 10$^{6}$ years.
The last Sun's position in the galactic equatorial plane occured before 3.87 $\times$ 10$^{6}$ years.
The last extinctions had to occur before (3.87 $\times$ 10$^{6}$ $-$ $\varepsilon$ $\times$ $a^{3/2}$) years, where 
semi-major axis $a$ of the comet is given in astronomical units and 0 $<$ $\varepsilon$ $<$ 1.
The value of $\varepsilon$ describes the initial position of the comet on its orbit ($\varepsilon$ $=$ 1/2 corresponds to aphelion).
Since the Solar System's oscilations are periodic, other possibilities are  [(n $\times$ 72.6 / 2 $+$ 3.87 ) $\times$ 10$^{6}$ $-$ $\varepsilon$ $\times$ $a^{3/2}$ ] years, 
where $n$ is an integer number, which multiplies the half period of the Solar System's oscilations. 
Thus, also (40.2 $\times$ 10$^{6}$  $-$  $\varepsilon$ $\times$ $a^{3/2}$) years
($n$ $=$ 1), or,  (76.5 $\times$ 10$^{6}$  $-$  $\varepsilon$ $\times$ $a^{3/2}$) years ($n$ $=$ 2) in the past.
The values could correspond to the value of about 65 $\times$ 10$^{6}$ years of the late Cretaceous extinctions which included the disappearance 
of the dinosaurs (Rampino and Stothers 1984). The mass extictions on the Earth have periodic behaviour with the period approximately 33 $\times$ 10$^{6}$ years.
The case $n$ $=$ 2 could occur for
$a$ $=$ 5.1 $\times$ 10$^{4}$ AU / $\varepsilon^{2/3}$.
The comet was increasing its heliocentric distance for some part
of its motion before the hit of the Earth, if
5.1 $\times$ 10$^{4}$ AU  $<$ $a$ $<$ 8.1 $\times$ 10$^{4}$ AU.
If the comet was still moving toward the Sun, then $\varepsilon$ $<$ 1/2
and $a$ $>$ 8.1 $\times$ 10$^{4}$ AU.
The case $n$ $=$ 1, for 33 $\times$ 10$^{6}$ years in the past, yields
$a$ $=$ 3.73 $\times$ 10$^{4}$ AU / $\varepsilon^{3/2}$.
If $\varepsilon$ $<$ 1/2, then $a$ $>$ 5.9 $\times$ 10$^{4}$ AU.
Similarly, the case $n$ $=$ 3, for 99 $\times$ 10$^{6}$ years in the past, yields
$a$ $=$ 5.74 $\times$ 10$^{4}$ AU / $\varepsilon^{3/2}$.
Although the galactic tidal forces exclude stability of the Oort cloud for
large values of semi-major axis, $a$ $>$ 8 $\times$ 10$^{4}$ AU,
a close encounter of a star or an interstellar cloud could disturb
a comet of the Oort cloud. Finally, there is also a possibility that objects
of interstellar origin hit the Earth.

There is an another feature of the effect of Solar System's vertical motion above and below the galactic plane 
on the possible terrestrial mass extinctions. We should await, on the basis of Fig. 7 and Table 3, 
that the period of possible exceptionally strong mass extinction is that
the period of possible mass extinctions is (4.5-5.0) $\times$ 10$^{8}$ years, if the galactic tide
is the reason of the mass extinctions.
This is twice the value presented by Schwartz and James (1984). It seems that we
are not able to explain the major extinction events on the basis of the galactic tide alone.

\section{Conclusion}
The paper treats the effect of the galactic tide on a cometary motion
with respect to the Sun. Our detailed numerical calculations show that
the $\Gamma-$terms play an important role in cometary orbital evolution.
The $\Gamma-$terms cause that periods of secular evolution of the orbital 
elements differ from the periods of conventional models. The real period
is 1.7-times smaller  than the period of the conventional models, if
the initial cometary inclination is close to 90 degrees, eccentricity is close to zero
and semi-major axis is about 5 $\times$ 10$^{4}$ AU. Shorter periods of secular evolution of orbital elements lead to
the conclusion that frequency of cometary returns into the inner part of
the Solar System is about two times higher than the frequency following from the conventional models.

A comet from the Oort cloud can be a source of the dinosaurs extinction at about 65 Myr ago. 
Either a close encounter of a star or an interstellar cloud  
disturbed a comet in the Oort cloud in the way that its semi-major axis increased/decreased above the value  
5.1 $\times$ 10$^{4}$ AU and the comet hit the Earth. Or,  an object of interstellar origin hit the Earth.

\section*{Acknowledgement}
This work was supported by the Scientific Grant Agency VEGA, Slovak Republic,
grant No. 2/0016/09 and by the Comenius University grant UK/405/2009.

\end{document}